\documentclass[pdflatex,sn-mathphys-ay]{sn-jnl}
\setcitestyle{authoryear,open={(},close={)},aysep={ },yysep={;},notesep={; }}

\usepackage{graphicx}%
\usepackage{multirow}%
\usepackage{amsmath,amssymb,amsfonts}%
\usepackage{amsthm}%
\usepackage{mathrsfs}%
\usepackage[title]{appendix}%
\usepackage{xcolor}%
\usepackage{textcomp}%
\usepackage{manyfoot}%
\usepackage{booktabs}%
\usepackage{algorithm}%
\usepackage{algorithmicx}%
\usepackage{algpseudocode}%
\usepackage{listings}%
\usepackage{siunitx}%
\usepackage{float}%
\usepackage{rotating}%
\DeclareSIUnit{\liter}{L}
\DeclareSIUnit{\rpm}{rpm}

\theoremstyle{thmstyleone}%
\theoremstyle{thmstyletwo}%

\theoremstyle{thmstylethree}%

\begin{document}

\title[Article Title]{Quantitative interpretation of Brookfield DV3TLV measurements: shear rate conversion, correction factors, and applicability limits}


\author*[1]{\fnm{Andrey E.} \sur{Vasiliev}}\email{vasiliev.a.e@hydro.nsc.ru}

\author[1]{\fnm{Alexey S.} \sur{Besov}}\email{besov.a.s@hydro.nsc.ru}

\author[1]{\fnm{Danil O.} \sur{Andreev}}\email{d.andreev1@g.nsu.ru}

\affil*[1]{\orgname{Lavrentyev Institute of Hydrodynamics of the Siberian Branch of the RAS}, \orgaddress{\street{Ac. Lavrentieva ave. 15}, \city{Novosibirsk}, \postcode{630090}, \country{Russian Federation}}}



\abstract{The flow behavior and hydrodynamic characteristics of fluids in rotational viscometry systems are investigated using the Brookfield DV3TLV viscometer, with emphasis on measurement reliability and applicability limits of different measuring geometries. The results are compared and validated using the high-precision MCR 302 rheometer manufactured by the Austrian company Anton Paar. Both Newtonian (water and glycerol) and non-Newtonian fluids (guar-based gels), exhibiting fundamentally different viscosity–shear rate behavior, were included in the study. Based on the comparison of measurements obtained with the Brookfield DV3TLV viscometer and the MCR 302 rheometer, empirical coefficients were determined that relate the spindle rotational speed to the shear rate, taking into account the geometry of the measuring systems. Analysis of the Reynolds number range showed that laminar flow conditions were maintained for all measurement systems, which justifies the application of quasi-static models that neglect possible flow turbulence within them. Comparison with high-precision measurements performed on the MCR 302 rheometer showed that, with appropriate interpretation, the data obtained using the Brookfield instrument can be used to estimate the real viscosity of process fluids with an accuracy specific to each geometry and its operating conditions. The proposed methodology enables reliable characterization of flow properties in rotational systems and can be applied in engineering practice and laboratory analysis of complex fluids, especially at oil and food production facilities where high-end rheometers are unavailable or impractical to use. The study is formulated within the framework of experimental fluid mechanics and non-Newtonian flow characterization.}




\maketitle

\section{Introduction}\label{sec1}

{Accurate and reproducible characterization of flow behavior, hydrodynamic properties and rheological properties of process fluids is required in many engineering applications, particularly in confined rotational flows. These properties govern the stability and control of many technological processes \citep{bib7}. One of the key characteristics measured using viscometers is viscosity, and among the most widely used instruments of this type are Brookfield viscometers, which have gained widespread adoption due to their relatively low cost, ease of operation, and minimal requirements for operator qualification. These instruments are widely used in the food, pharmaceutical, cosmetic, chemical and petroleum refining industries \citep{bib2,bib3}; however, questions remain regarding the reliability of the obtained results and the conditions (ranges) of applicability of the measurement systems supplied with these viscometers. Some researchers even use the term “Brookfield viscosity,” which reflects concerns regarding the accuracy of measurements obtained with these instruments \citep{bib7}.}

{Despite their widespread use and several technical advantages over other instruments of this type, Brookfield viscometers do not allow direct measurement of absolute values of shear viscosity. Instead, the instrument reports the so-called “apparent viscosity” (or “Brookfield viscosity”), a conditional parameter calculated by the instrument software \citep{bib7}. This is due to the absence of a rigorous analytical relationship between torque and shear rate for the DV3TLV model. Previously, considerable effort was devoted by \cite{bib16} to developing a rigorous analytical model for Brookfield R.V. viscometers, with further consideration by \cite{bib11} to employ a zero-thickness disk model for disk-shaped spindles in calculations. However, the conversion factors included in the manufacturer's official technical documentation, which engineers could use for routine measurements, are not provided for all configurations and types of fluid \citep{bib8,bib9}.}

{The geometry of the spindles used (disk, cylindrical and spherical) does not ensure a uniform shear flow within the measured volume \citep{bib7}. This circumstance limits the possibility of direct interpretation of the results obtained, and the literature emphasizes the need to introduce correction factors to improve the accuracy and reproducibility of viscosity measurements performed with these instruments \citep{bib13,bib17}.}

{It is possible that the Brookfield DV3TLV viscometer measurement systems used in this study -- namely the LV1, LV2, LV3, LV4 spindles and the UL Adapter -- also generate non-uniform flow within the samples and may therefore provide inaccurate information about the rheological properties of the fluid within the investigated shear rate range.}

{The purpose of this study is to establish a physically consistent methodology for interpreting flow behavior in rotational viscometry systems using the Brookfield DV3TLV viscometer for the use of the full set of available setups. Although the manufacturer produces other modifications of this instrument with different design features and application areas, the DV3TLV model was selected due to its versatility, widespread use, relatively low cost, and convenience for routine measurements.}

{The viscosity value reported by the instrument is largely determined by the configuration and geometry of the measurement system, as well as by the temperature, sample volume and immersion depth of the spindle, which may lead to distortion of information about the actual rheological properties of the fluid investigated; these factors complicate the comparison of results obtained with Brookfield viscometers with data from higher-class rheometers \citep{bib8,bib13,bib17}. However, high-end rheometers cannot achieve such widespread use due to their high cost and demand for highly qualified personnel, whereas expanding the capabilities of Brookfield-type viscometers, such as the DV3TLV, through the development of methods to obtain reliable results with these instruments represents the only practical alternative.}

{Guided by similar considerations, many researchers have attempted to improve Brookfield measurement systems, determine correction coefficients, and develop methods to convert apparent viscosity into absolute viscosity values \citep{bib6,bib12,bib15}. For example, attempts have been made to employ specialized “cup–spindle” systems, and studies have been conducted to determine the viscoelastic properties of liquid samples based on the data obtained \citep{bib15}. As a result, it was shown that, with an appropriate selection of the spindle, temperature, and sample preparation procedure, the Brookfield viscometer can provide rotational flow measurements of non-Newtonian fluids exhibiting thixotropic behavior with sufficient repeatability and reproducibility, which is not achievable using standard methods with conventional spindle \citep{bib6}. Thus, \cite{bib15} proposed the use of the “Cup 1” with the Brookfield YR-I rheometer to measure the viscoelastic properties of liquids and developed recommendations for selecting methods to convert spindle rotational speed into shear rate when using a cup–spindle measurement system. The application of this methodology makes it possible to obtain results that agree with the data of more expensive high-precision rheometers with an error not exceeding 10\% \citep{bib15}. Moreover, as shown by \cite{bib12}, for materials exhibiting pronounced elastic properties and complex structural behavior of the fluid at different shear rates, a reliable evaluation of rheological parameters is possible only under controlled deformation conditions, which cannot be performed using the Brookfield DV3TLV viscometer and therefore makes attempts to adapt it for such rotational flow measurements impractical. At the same time, the literature contains relatively few systematic studies of viscosity characteristics obtained using different models of Brookfield viscometers, as well as investigations of various types of process fluids performed with these instruments. This complicates the development of universal recommendations for the use of Brookfield viscometers and raises concerns regarding the reliability of the results obtained from measurements performed with these instruments. In addition, it is not always possible to modify standard geometries, which limits the range of tasks that can be addressed and, consequently, the number of testing laboratories using Brookfield viscometers.}

{Moreover, the interpretation of the data obtained using Brookfield viscometers is further complicated by the need to determine and account for the fluid flow regime in each case. For this purpose, the Reynolds number ($Re$) is used — a dimensionless criterion that allows the flow to be classified as laminar, transitional, or turbulent \citep{bib7}. Elevated Reynolds number values indicate the possibility of flow turbulence in the space of operation of the Brookfield spindles, which may potentially account for discrepancies between results obtained using the Brookfield viscometer and those measured with higher-class rheometers such as the Anton Paar MCR 302. Accounting for such effects is necessary for the correct interpretation of the rheological data obtained with the Brookfield viscometer, which was implemented in the course of this study.}


{The main contributions of this work are as follows. First, analytical relationships are derived to convert the spindle rotation speed into the shear rate for the LV2 and LV3 spindles, for which such relations are not provided in the manufacturer’s documentation. Second, empirical correction coefficients are determined based on direct comparison with high-precision rheometer measurements, allowing the conversion of apparent Brookfield viscosity into values approaching absolute viscosity. Third, the applicability limits of different geometries are systematically analyzed in viscosity–shear rate coordinates, with explicit consideration of the Reynolds number and flow regime transitions.}

{These results provide a unified framework for quantitative interpretation of Brookfield viscometer data and define the conditions under which such rotational flow measurements can be considered physically reliable.}

\section{Methods}\label{sec4}

\subsection{Materials and fluids}\label{subsec4}

{Both Newtonian and non-Newtonian fluids were investigated in this study. The first group consisted of distilled water and aqueous glycerin solutions with concentrations of 92.5\% and 50\%, used as reference fluids with well-characterized rheological properties. Non-Newtonian samples included guar gum solutions (WGA NG-1) with concentrations of 2.4 and 3.6~\si{\kilogram\per\meter\cubed} (linear gels), as well as a cross-linked guar gel with a concentration of 2.4~\si{\kilogram\per\meter\cubed}. Guar gum solutions were prepared by gradually introducing the thickener powder into the vortex formed by the rotating blades of a laboratory mixer in a beaker filled with distilled water. Mixing was carried out at room temperature (T $\approx$ 25~\si{\degreeCelsius}) until complete dissolution of the powder and disappearance of agglomerates, which typically required about 10 minutes. The mixer was operated at a rotational speed of 1500 rpm, after which the gel was allowed to rest for 30 minutes.}

\subsection{Instruments}\label{subsec4}

{The rheological properties of the fluids were measured using two instruments: a Brookfield DV3TLV rotational viscometer and an Anton Paar MCR 302 rheometer. The viscometer was equipped with LV1–LV4 spindles as well as a UL Adapter intended for low-viscosity measurements under temperature-controlled conditions. For temperature control of guar gum gel samples during measurements with the LV-series spindles, a \SI{600}{\milli\liter} jacketed beaker with inlet and outlet ports (drawing 2-1233-05, Khimlaborpribor) was used. The Anton Paar MCR 302 rheometer was used to obtain calibration viscosity curves over the widest possible range of shear rates. The Anton Paar CC27 measuring system, based on a concentric cylinder geometry, was used in this study.}

{Sample temperature during Brookfield measurements was controlled using an external circulating liquid thermostat (LOIP FT-211-25) with distilled water as the heat-transfer medium, both for the UL Adapter and for the jacketed beaker. Temperature was monitored throughout the experiment using a built-in PT1000 temperature probe and additionally checked with a digital thermometer (Hanna Instruments Checktemp HI 98501). For the MCR 302 rheometer, temperature control was achieved using the standard C-PTD200 temperature control system.}

{For each fluid sample tested with the Brookfield viscometer, the appropriate spindle was selected based on the sample viscosity and the available measurement range. If the initial sample temperature differed from the set temperature, the sample was allowed to equilibrate in the measuring cell before the measurements. The viscometer recorded data at intervals of 120 s for the first measurement point and 60 s for all subsequent points, and the viscosity at each shear rate was calculated as the average of five measurements.}

{Viscosity curves for the same samples were recorded on MCR 302 rheometer in shear-rate-controlled mode over the range 0.01–1000 \si{s^{-1}}. Temperature was maintained using the rheometer’s built-in Peltier system with an accuracy of 0.1~\si{\degreeCelsius}.}

\subsection{Geometry of measuring systems}\label{subsec4}

{The geometric parameters of the standard Brookfield spindles are summarized in Table~\ref{tab:spindles}.}

\begin{sidewaystable}
\caption{Geometric parameters of Brookfield spindles}
\label{tab:spindles}
\centering
\begin{tabular}{lcccc}
\hline
\textbf{Spindle} & \textbf{Cylinder diameter, mm} & \textbf{Effective length\footnotemark[1], mm} & \textbf{Rod diameter, mm} & \textbf{Height to disk, mm} \\
\hline
LV1 (\#61) & 18.842 & 74.93 & 3.2 & 80.97 \\
LV4 (\#64) & 3.176 & 33.96 & 3.2 & --- \\
\hline
\multicolumn{5}{c}{} \\
\hline
\textbf{Spindle} & \textbf{Disk diameter, mm} & \textbf{Disk height, mm} & \textbf{Rod diameter, mm} & \textbf{Height to disk, mm} \\
\hline
LV2 (\#62) & 18.72 & 25.40 & 3.2 & 50.00 \\
LV3 (\#63) & 12.60 & 25.60 & 3.2 & 50.00 \\
\hline
\multicolumn{5}{c}{} \\
\hline
\textbf{Spindle} & \textbf{Inner cup diameter, mm} & \textbf{Effective length\footnotemark[1], mm} & \textbf{Cylinder diameter, mm} & \textbf{Cup height, mm} \\
\hline
UL Adapter (UL) & 27.62 & 92.39 & 25.13 & 90.74 \\
\hline
\end{tabular}
\footnotetext[1]{The effective length includes an end-effect correction and should be used in the calculation equations given in \cite{bib3}}
\end{sidewaystable}

{For LV spindles, the correct liquid volume was determined using the mark on the spindle shaft in accordance with the manufacturer's documentation. For the UL Adapter, a fixed liquid volume of \SI{16}{\milli\liter} was used, as required by the operating instructions for complete filling of the measuring chamber.}

{Measurements on the Anton Paar rheometer were performed using a CC27 concentric cylinder geometry with a bob diameter of 27 mm and a gap of 1.13 mm between the bob and the cup.}

\subsection{Measurement procedure}\label{subsec4}

{Viscosity measurements on the Brookfield DV3TLV viscometer were carried out at room temperature and at 20, 30, and 70~\si{\degreeCelsius} over a spindle speed range of 0.1–250 rpm. The angular velocity $\omega$ was calculated according to the formula given in \cite{bib2,bib3}.}

\begin{equation}
\omega = \frac{2\pi N}{60} \quad [\mathrm{rad\cdot s^{-1}}]
\label{eq:omega}
\end{equation}

{where $N$ is the spindle rotation speed, rpm.}

{For the Brookfield viscometer, torque, spindle rotational speed, apparent viscosity (Brookfield viscosity), and temperature were recorded. The shear rate was calculated using manufacturer-provided coefficients for the LV1, LV4, and UL Adapter spindles and experimentally determined coefficients for the LV2 and LV3 spindles. For the LV2 and LV3 spindles, empirically determined conversion coefficients $k$ of 0.50 and 0.46, respectively, obtained in the present study, were used.}

{Measurements of dynamic viscosity and shear rate control on the Anton Paar MCR 302 rheometer were carried out automatically using the RheoCompass software.}

\subsection{Reynolds number calculation}\label{subsec4}

The Reynolds number ($Re$) was calculated to determine the flow regime (laminar, transitional, or turbulent) in the gap between the rotating spindle and the stationary cup according to the following relation:

\begin{equation}
\mathrm{Re} = \frac{\omega \cdot \rho \cdot \left( R_e^2 - R_i^2 \right)}{2 \cdot \eta}
\label{eq:Re}
\end{equation}

{where $\omega$ is the angular velocity, \si{\radian\per\second} (calculated as $2\pi N/60$, where $N$ is the spindle rotational speed, \si{\rpm}); $\rho$ is the fluid density, \si{\kilogram\per\meter\cubed}; $R_e$ is the radius of the outer cylinder (cup), \si{\meter}; $R_i$ is the radius of the inner cylinder (spindle), \si{\meter}; and $\eta$ is the dynamic viscosity of the fluid, \si{\pascal\second}.}

{The calculations were performed separately for each measuring system, including the LV2 and LV3 spindles and the UL Adapter, taking into account the corresponding radii and rotational speed ranges. The Reynolds number was used to classify the flow regime as laminar ($Re$ $<$ 1), transitional (1 $\leq$ Re $<$ 1000), where the onset of end effects may occur, or turbulent ($Re$ $\geq$ 1000), in accordance with the approach described in \cite{bib7}.}

{The experimental results showed that rotational flow measurements of distilled water using the UL Adapter were carried out at Reynolds numbers in the range 95 $\leq$ Re $\leq$ 183, indicating the occurrence of end effects \citep{bib7}. For 92.5\% glycerol, the Reynolds number remained below Re $<$ 1, indicating a laminar flow regime. For 50\% glycerol, the Reynolds number ranged from 3 to 50, with viscosity values between 4 and 7~\si{\milli\pascal\second}, also indicating the presence of end effects. For guar gum–based fluids, the Reynolds number remained below $Re$ $<$ 2, indicating a laminar flow regime during the measurements with only minor end effects.}

\subsection{Data processing and comparison}\label{subsec4}

{For comparison of the results obtained with the Brookfield viscometer and the Anton Paar rheometer, identical shear rate values were selected for each experimental point. Data processing was performed in Microsoft Excel. If differences in shear rate occurred because the Brookfield viscometer sets the spindle speed in rpm, the measured data were interpolated over the corresponding interval using cubic splines in OriginPro, allowing comparable viscosity curves to be constructed and deviations to be evaluated. The reproducibility, mean value, and standard deviation were evaluated for repeated measurements. Differences between the instruments were assessed using absolute and relative viscosity deviations at identical or similar shear rates.}

\subsection{Measurement limits}\label{subsec4}

{For the Brookfield viscometer, the recommended torque range was taken into account, and only measurements within 10--90\% of the maximum torque were considered reliable. 
According to the manufacturer’s documentation \citep{bib2,bib3}, turbulent effects and sample heating may occur at rotational speeds above 100 rpm, particularly when measurements are performed in an open cup without temperature control. The influence of turbulent effects was considered during data interpretation through calculation of the Reynolds number and identification of the corresponding flow regime. Sample overheating when using an open cup was prevented by using a LOIP FT-211-25 cryothermostat with continuous temperature monitoring by the Brookfield viscometer thermocouple throughout the experiments.}

\subsection{Correction coefficient}\label{subsec4}

{To improve the reliability of rotational flow measurements of technological fluids, a correction coefficient $k_{corr}$ was introduced to convert the Brookfield apparent viscosity to values approaching the absolute viscosity. The correction coefficient was calculated as the arithmetic mean of all measurement points:}

\begin{equation}
k_{\mathrm{corr}} = \frac{1}{n} \sum_{i=1}^{n} \left| \frac{\eta_{A,i}}{\eta_{B,i}} \right|
\label{eq:kcorr}
\end{equation}

{where $i$ is the index of the measurement point at a matching (or adjusted) shear rate; $n$ is the number of viscosity measurements at different shear rates; $\eta_{A,i}$ is the dynamic viscosity measured with the Anton Paar rheometer at point $i$; and $\eta_{B,i}$ is the apparent viscosity measured with the Brookfield viscometer at point $i$.}

{The corrected Brookfield viscosity was calculated using the following expression:}

\begin{equation}
\eta_{corrB,i} = k_{\mathrm{corr}} \cdot \eta_{B,i}
\quad [\mathrm{Pa\cdot s}]
\label{eq:eta_corr}
\end{equation}

\section{Shear rate calculation for LV2 and LV3 spindles}\label{sec3}

{For proper hydrodynamic interpretation of data obtained with the Brookfield DV3TLV viscometer, the spindle rotational speed (rpm) should be converted into a physically meaningful parameter --- the average shear rate $\dot{\gamma}$ --- taking into account the geometry of the specific spindle.}

\begin{figure}[H]
\centering
\includegraphics[width=\columnwidth,height=0.8\textheight,keepaspectratio]{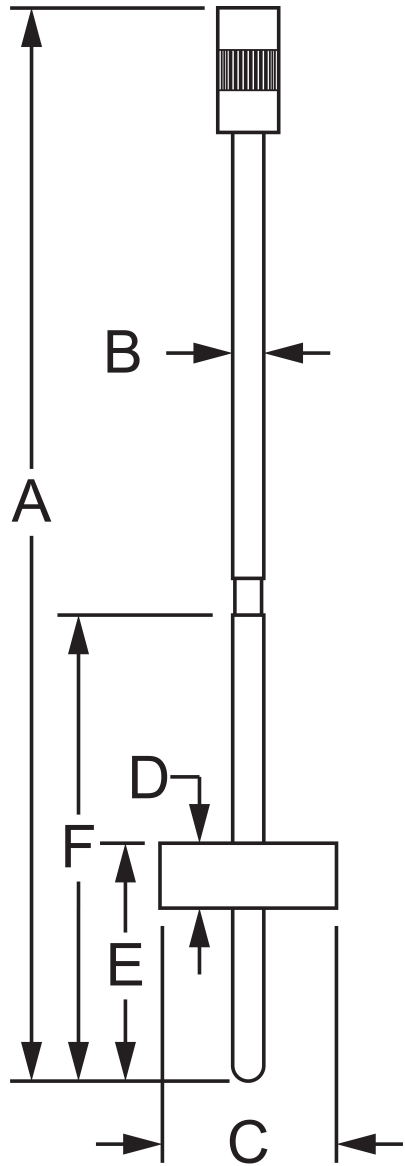}
\caption{Schematic representation of the LV2 and LV3 spindles. The geometry consists of three coaxial sections: a cylindrical shaft (B), a disk-shaped cylindrical enlargement (D), and a lower rod immersed in the fluid (E). The characteristic dimensions include the shaft length (A), immersion depth (F), and cup diameter (C). Each section contributes to the shear flow generated in the annular gap between the rotating spindle surface and the stationary cup wall.}
\label{fig:lv2_lv3_geometry}
\end{figure}

{The LV2 and LV3 spindles are axisymmetric bodies consisting of three coaxial sections: a disk-shaped cylindrical enlargement, a cylindrical shaft connecting the spindle to the viscometer, and a rod located below the disk (Fig.~\ref{fig:lv2_lv3_geometry}). Each of these sections contributes to the overall shear flow developed in the gap between the rotating surface of the spindle and the wall of the stationary cup.}

{The Brookfield DV3TLV viscometer records torque in relative units, expressed as a percentage of the instrument’s full-scale value (Torque, \%). For subsequent calculations, this value must first be converted into physical units (\si{\milli\newton\meter}), after which the corresponding torque can be related to the shear stress at the spindle surface and then to the shear rate.}

{The full-scale torque $T_{full scale}$ can be expressed as:}

{\[
T_{\mathrm{full\ scale}} = 6.737 \cdot 10^{-5}~\si{\newton\meter}
\]}

{The actual torque can be calculated as:}

\begin{equation}
T_{\mathrm{LV}} = \frac{T_{\mathrm{full\ scale}} \cdot Torque}{100\%} \quad [\si{\newton\meter}]
\label{eq:torque}
\end{equation}

{This value is taken from the Brookfield documentation (DV3T Instructions) for each specific spindle and converted to SI units \citep{bib2}. The expression for the local shear rate in the annular gap is given in the classical rheology textbook by \cite{bib7} and further detailed in the Brookfield practical rheology guidelines \citep{bib3}. The geometric description of the spindles used and the design of the measuring system follow the operating recommendations provided in the Brookfield documentation \citep{bib7}.}

\subsection{Spindle geometry}\label{subsec3}

{Each spindle (LV2 and LV3) consists of three coaxial cylindrical sections: the shaft ($v$), the rod ($s$), and the disk ($d$). The geometric parameters are given in meters Table~\ref{tab:geometry}.}

\begin{table}[h]
\caption{Dimensions of the spindles}
\label{tab:geometry}
\centering
\begin{tabular}{lccc}
\hline
Parameter & LV2 (\#62) & LV3 (\#63) \\
\hline
Shaft radius ($R_v$) & 0.00160 & 0.00160 \\
Shaft length ($L_v$) & 0.02460 & 0.02440 \\
Rod radius ($R_s$) & 0.00160 & 0.00160 \\
Rod length ($L_s$) & 0.01854 & 0.02382 \\
Disk radius ($R_d$) & 0.00936 & 0.00630 \\
Disk length ($L_d$) & 0.00866 & 0.00178 \\
\hline
\end{tabular}
\end{table}

{The geometric coefficient is given by:}

\begin{equation}
K_{\tau} = 2\pi \left( R_v^2 \cdot L_v + R_s^2 \cdot L_s + R_d^2 \cdot L_d \right) [\si{\meter\cubed}]
\label{eq:Ktau}
\end{equation}

Substitution yields:

\begin{gather*}
K_{\tau(\mathrm{LV2})} = 4.470 \cdot 10^{-6}~\si{\meter\cubed} \\
K_{\tau(\mathrm{LV3})} = 1.220 \cdot 10^{-6}~\si{\meter\cubed}
\end{gather*}

\subsection{Shear stress calculation}\label{subsec3}

{The shear stress is given by:}

\begin{equation}
\tau = \frac{T}{K_{\tau}} \quad [\si{\pascal}]
\label{eq:tau}
\end{equation}

{After substituting the torque values (in \si{\newton\meter}) and the corresponding coefficient ${K_{\tau}}$, the shear stress $\tau$ is obtained in \si{\pascal}.}

\subsection{Relation between viscosity and shear rate}\label{subsec3}

{The shear rate is given by:}

\begin{equation}
\dot{\gamma} = \frac{\tau}{\eta} \quad [\si{\per\second}]
\label{eq:shear_rate}
\end{equation}

{where $\eta$ is the dynamic viscosity obtained from the viscometer readings, \si{\pascal\second}, and $\tau$ is the shear stress calculated from the torque.}

{The shear rate $\dot{\gamma}$ was divided by the spindle rotational speed $N$ (rpm) to obtain the conversion coefficient:}

\begin{equation}
k = \frac{\dot{\gamma}}{N} \quad [\si{\per\second\per\rpm}]
\label{eq:k}
\end{equation}

{The values of $k$ were calculated for each measurement. The resulting data sets were averaged after removing outliers.}

\subsection{Statistical analysis}\label{subsec3}

{The standard deviation was calculated using the following expression:}

\begin{equation}
\sigma = \sqrt{\frac{1}{n - 1} \sum_{i=1}^{n} \left( k_i - \bar{k} \right)^2}
\label{eq:sigma}
\end{equation}

{where $k_i$ are the individual coefficients, $\bar{k}$ is the mean value, and the dataset included all points satisfying the following criteria: torque within the range of 10--90\% and viscosity measurement accuracy within acceptable limits.}

\subsection{Obtained conversion coefficients}\label{subsec3}

{A larger set of experimental results than those presented in this paper was used to calculate the coefficients, including additional measurements of the previously mentioned fluids and a based on PT GPS polyacrylamide (PAM) solution with a concentration of 6~\si{\kilogram\per\meter\cubed}.}

\begin{table}[htbp]
\caption{Calculated conversion coefficients $k$ for LV2 spindle}
\label{tab:k_LV2}
\centering
\begin{tabular}{lccccc}
\hline
Fluid & Conc. & Temp.,~\si{\degreeCelsius} & Mean $k$ & $\sigma$ & Points \\
\hline
Glycerol & 92.5\% & 25.7 & 0.500218 & 0.000266 & 10 \\
Glycerol & 92.5\% & 20 & 0.500547 & 0.000278 & 4 \\
Linear guar & 2.4~\si{\kilogram\per\meter\cubed} & 28.7 & 0.500744 & 0.000106 & 8 \\
Linear guar & 3.6~\si{\kilogram\per\meter\cubed} & 27.5 & 0.500218 & 0.000266 & 10 \\
Linear guar & 2.4~\si{\kilogram\per\meter\cubed} & 20 & 0.501812 & 0.000883 & 6 \\
Linear guar & 2.4~\si{\kilogram\per\meter\cubed} & 30 & 0.502557 & 0.000603 & 7 \\
Linear guar & 3.6~\si{\kilogram\per\meter\cubed} & 20 & 0.502342 & 0.000307 & 8 \\
Linear guar & 3.6~\si{\kilogram\per\meter\cubed} & 30 & 0.502387 & 0.000500 & 8 \\
Cross-linked guar & 2.4~\si{\kilogram\per\meter\cubed} & 70 & 0.502430 & 0.000065 & 5 \\
\hline
Average & --- & --- & 0.50124 & 0.00085 & 66 \\
\hline
\end{tabular}
\end{table}

\begin{table}[!htbp]
\caption{Calculated conversion coefficients $k$ for LV3 spindle}
\label{tab:k_LV3}
\centering
\begin{tabular}{lccccc}
\hline
Fluid & Conc. & Temp.,~\si{\degreeCelsius} & Mean $k$ & $\sigma$ & Points \\
\hline
Glycerol & 92.5\% & 26 & 0.460200 & 0.000260 & 9 \\
Glycerol & 92.5\% & 20 & 0.460070 & 0.000340 & 9 \\
Linear guar & 3.6~\si{\kilogram\per\meter\cubed} & 27.1 & 0.460051 & 0.000038 & 4 \\
Linear guar & 3.6~\si{\kilogram\per\meter\cubed} & 20 & 0.460074 & 0.000842 & 7 \\
Linear guar & 3.6~\si{\kilogram\per\meter\cubed} & 30 & 0.460520 & 0.000500 & 5 \\
Cross-linked guar & 2.4~\si{\kilogram\per\meter\cubed} & 20 & 0.460233 & 0.000217 & 10 \\
Cross-linked guar & 2.4~\si{\kilogram\per\meter\cubed} & 70 & 0.460162 & 0.000172 & 2 \\
PAM & 6~\si{\kilogram\per\meter\cubed} & 70 & 0.460060 & 0.000040 & 4 \\
PAM & 6~\si{\kilogram\per\meter\cubed} & 70 & 0.464090 & 0.004300 & 20 \\
PAM & 6~\si{\kilogram\per\meter\cubed} & 70 & 0.462200 & 0.002360 & 20 \\
\hline
Average & --- & --- & 0.46019 & 0.00120 & 90 \\
\hline
\end{tabular}
\end{table}

{The tables present the calculated values of the coefficient $k$, relating the spindle rotational speed $N$ (rpm) to the average shear rate $\dot{\gamma}$ (\si{s^{-1}}), both for individual tests and for the combined dataset for each spindle Table~\ref{tab:k_LV2}, \ref{tab:k_LV3}. The coefficients show high stability: the standard deviation $\sigma$ for most fluids is below 0.0005 (in relative units), corresponding to a relative error of less than 0.1\%.}

{For the LV2 spindle, all calculated values of $k$ fall within a narrow range from 0.5002 to 0.5024. Even for the guar gel with a concentration of 2.4~\si{\kilogram\per\meter\cubed}, which exhibits pronounced non-Newtonian behavior, the deviation remains within acceptable limits and does not exceed 0.001. This confirms that the coefficient is stable with respect to variations in the rheological properties of the fluid and reflects its geometric origin.}

{For the LV3 spindle, the range of mean values is wider (0.46006–0.46409), which can be attributed to the higher sensitivity of the results to temperature, as temperature-dependent tests were included, as well as to measurement conditions at high viscosities and high shear rates. The largest standard deviation was observed for the tests conducted after pumping the PAM solution through a proppant pack, which may be attributed to structural instability of the solution or local polymer degradation. However, even in this case, the coefficient remains close to the overall mean value (0.46207), indicating the reproducibility of the calculations.}

{Thus, the obtained conversion coefficients can be considered universal for the LV2 and LV3 spindles under conditions similar to those of the present experiments (low torque range, stable Newtonian or non-Newtonian fluids, and no pronounced time-dependent structural instability). The mean values of $k$ are:}

\begin{gather*}
k_{\mathrm{LV2}} = 0.50124 \pm 0.00085 \quad [\mathrm{s^{-1}\cdot rpm^{-1}}] \\
k_{\mathrm{LV3}} = 0.46019 \pm 0.00120 \quad [\mathrm{s^{-1}\cdot rpm^{-1}}]
\end{gather*}

{These results can be used to construct generalized viscosity curves and to interpret data obtained with the Brookfield DV3TLV viscometer on a unified shear-rate scale using the expression given in \cite{bib3}:}

\begin{equation}
\dot{\gamma} = k \cdot N \quad [\mathrm{s^{-1}}]
\label{eq:shear_rate_k}
\end{equation}

{The procedure for converting torque to shear stress and subsequently to shear rate allows numerical values of the rpm--$\dot{\gamma}$ conversion coefficients to be obtained. The obtained values demonstrate high stability for both Newtonian fluids and non-Newtonian systems.}

\section{Results}\label{sec2}


\subsection{Comparison of Brookfield and Anton Paar measurements}\label{subsec2}

{Measurement results for distilled water, glycerol solutions, linear and cross-linked guar gels, are presented and compared using the Ametek Brookfield DV3TLV viscometer and the Anton Paar MCR 302 rheometer. The main objective of this study was to evaluate the precision of the viscosity measurements of process fluids performed using the Brookfield viscometer at room and reservoir temperatures, to obtain correction coefficients $k_{corr}$ to convert the Brookfield measurement results into absolute values, and to develop recommendations to improve the accuracy of these measurements.}

{The correction coefficient was calculated by dividing the viscosity values measured with the Anton Paar rheometer at each shear rate by the viscosity values measured with the Brookfield viscometer at the same shear rate. Subsequently, the arithmetic mean of all measurement results was calculated for each spindle.}

\begin{sidewaystable}
\caption{Comparison of viscosity measurements obtained with the Brookfield viscometer using different spindle geometries (LV1–LV4, UL Adapter) and the Anton Paar rheometer with the CC27 concentric cylinder system}
\label{tab:kcorr_all}
\small
\renewcommand{\arraystretch}{1.15}

\begin{tabular*}{\textheight}{@{\extracolsep\fill}cccccccc}
\toprule
Type & Fluid & Concentration & Temperature, \si{\degreeCelsius} & Spindle &
\multicolumn{2}{c}{Discrepancy, \%} & Mean $k_{\mathrm{corr}}$ \\
\cmidrule(lr){6-7}
& & & & & Range & Average & \\
\midrule

\multirow{7}{*}{Newtonian}
& \multirow{6}{*}{Glycerol}
& \multirow{5}{*}{92.5\%}
& \multirow{5}{*}{20}
& ULA & 3.7--5.0 & 4.4 & 1.1 \\
& & & & LV1 & 15.6--16.1 & 15.8 & 0.8 \\
& & & & LV2 & 7.6--8.5 & 8.3 & 0.9 \\
& & & & LV3 & 3.8--4.2 & 4.0 & 1.0 \\
& & & & LV4 & 18.5--26.3 & 21.9 & 0.8 \\
& & 50\% & 20 & ULA & 10.1--17.7 & 12.5 & 1.1 \\
\cmidrule(lr){2-8}

& Water & --- & 20 & ULA & 7.4--8.3 & 8.0 & 0.9 \\
\midrule

\multirow{23}{*}{Non-Newtonian}
& \multirow{16}{*}{Linear guar gel}
& \multirow{6}{*}{2.4~\si{\kilogram\per\meter\cubed}}
& \multirow{3}{*}{20}
& ULA & 1.9--3.4 & 2.8 & 1.0 \\
& & & & LV1 & 25.3--30.4 & 27.5 & 0.7 \\
& & & & LV2 & 43.9--53.7 & 48.6 & 0.5 \\
& & & \multirow{3}{*}{30}
& ULA & 0.4--5.7 & 2.3 & 1.0 \\
& & & & LV1 & 23.3--31.8 & 27.3 & 0.7 \\
& & & & LV2 & 43.3--55.5 & 49.2 & 0.5 \\
\cmidrule(lr){3-8}

& & \multirow{10}{*}{3.6~\si{\kilogram\per\meter\cubed}}
& \multirow{4}{*}{20}
& ULA & 7.2--14.2 & 10.0 & 1.1 \\
& & & & LV1 & 31.3--33.7 & 32.8 & 0.7 \\
& & & & LV2 & 48.1--52.5 & 50.9 & 0.5 \\
& & & & LV3 & 55.9--56.3 & 56.1 & 0.4 \\
& & & \multirow{4}{*}{30}
& ULA & 0.5--2.4 & 1.0 & 1.0 \\
& & & & LV1 & 27.5--29.5 & 28.6 & 0.7 \\
& & & & LV2 & 44.7--51.4 & 48.6 & 0.5 \\
& & & & LV3 & 55.2--55.4 & 55.3 & 0.4 \\
\cmidrule(lr){2-8}

& \multirow{7}{*}{Cross-linked guar gel}
& \multirow{7}{*}{2.4~\si{\kilogram\per\meter\cubed}}
& \multirow{5}{*}{70}
& ULA & 0.3--9.0 & 3.5 & 1.0 \\
& & & & LV1 & 2.6--6.2 & 4.3 & 1.0 \\
& & & & LV2 & 127.4--275.2 & 225.0 & 3.3 \\
& & & & LV3 & 10.0--149.7 & 20.5 & 1.1 \\
& & & & LV4 & 20.0--57.6 & 43.6 & 0.6 \\
& & & \multirow{2}{*}{20}
& LV3 & 23.0--68.7 & 55.0 & 0.5 \\
& & & & LV4 & 21.0--64.0 & 41.2 & 0.6 \\
\botrule
\end{tabular*}
\end{sidewaystable}

\subsubsection{Results of Testing a 92.5\% Glycerol Solution}\label{subsubsec2}

\begin{figure}[H]
\centering
\includegraphics[width=\textwidth]{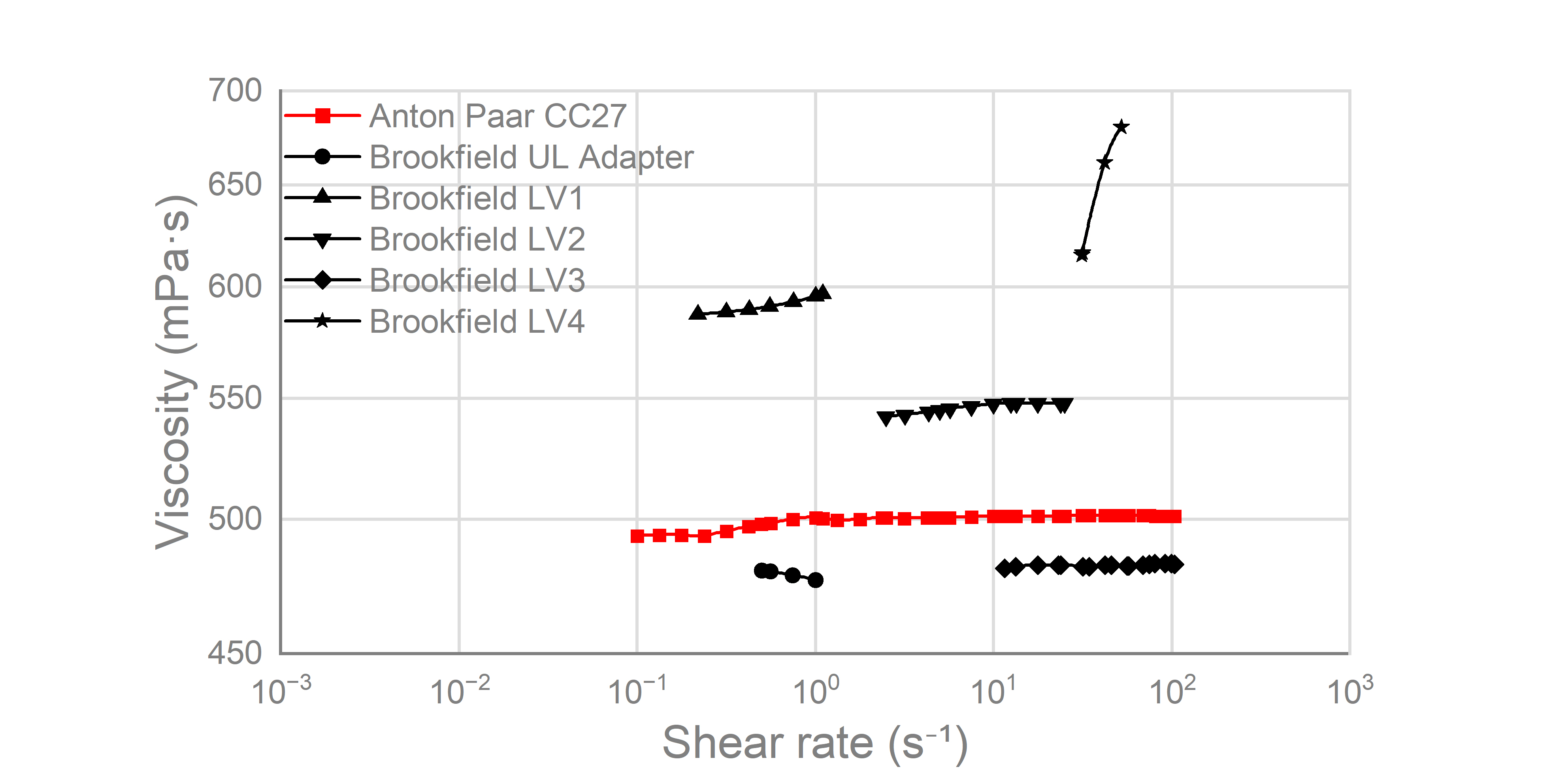}
\caption{Viscosity as a function of shear rate for a 92.5\% glycerol solution at 20~\si{\degreeCelsius}, measured using the Anton Paar rheometer (CC27) and the Brookfield viscometer with different spindle geometries (UL Adapter, LV1–LV4).}
\label{fig:glycerol_20C}
\end{figure}

\begin{figure}[H]
\centering
\includegraphics[width=\textwidth]{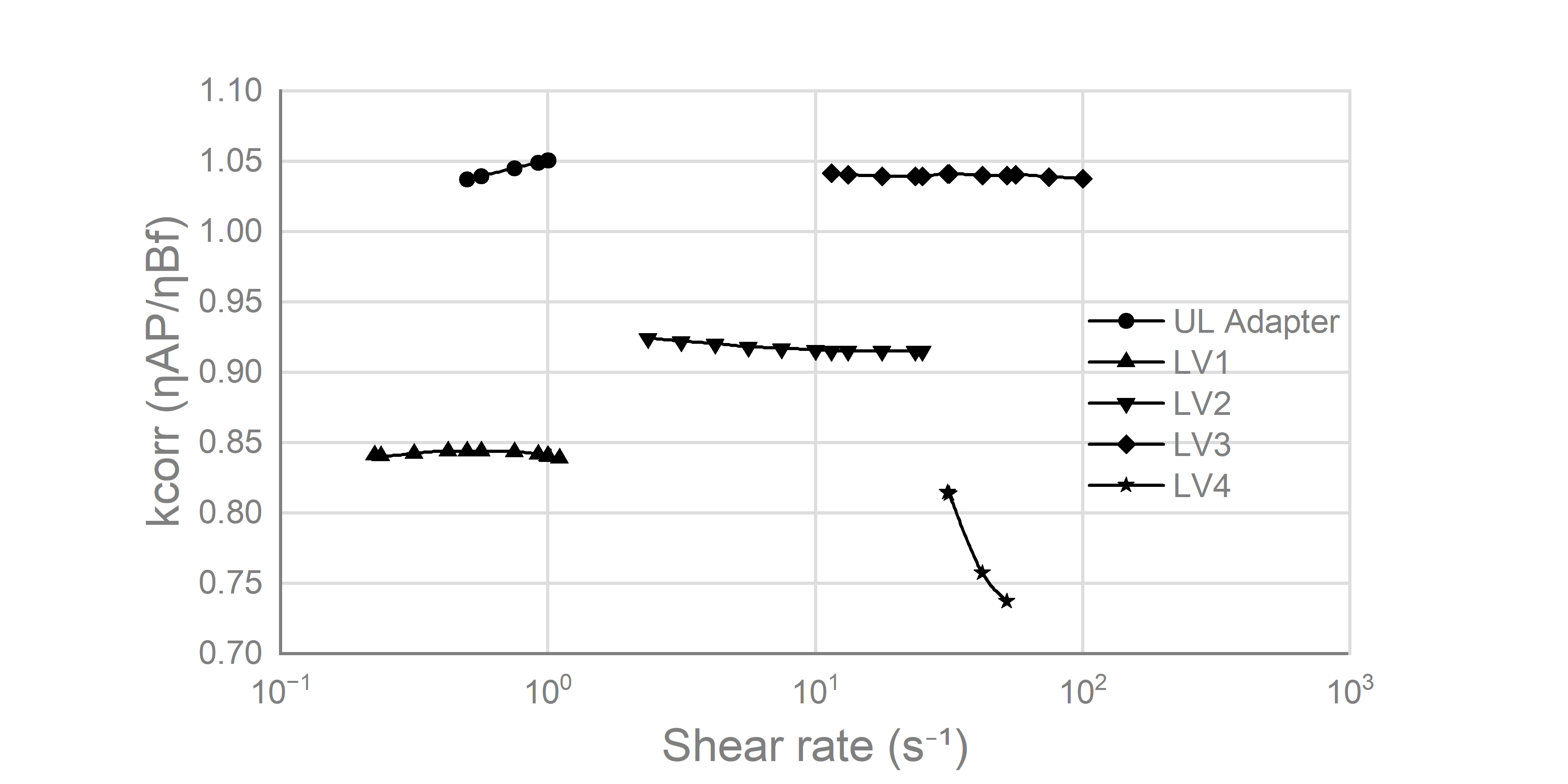}
\caption{Correction coefficient $k_{\mathrm{corr}} = \eta_{\mathrm{AP}} / \eta_{\mathrm{Bf}}$ as a function of shear rate for a 92.5\% glycerol solution at 20~\si{\degreeCelsius} obtained for different Brookfield spindle geometries.}
\label{fig:kcorr_glycerol}
\end{figure}

{For the 92.5\% glycerol solution, measurements were performed using the LV1–LV4 spindles and the UL Adapter system, with temperature control at T = \SI{20}{\degreeCelsius} (Fig.~\ref{fig:glycerol_20C}). The results closest to the absolute values were obtained when using the UL Adapter system and the LV3 spindle. A shear rate interval of $\dot{\gamma}$ from 1.1 to 2.5 \si{s^{-1}} was identified that could not be covered during viscosity measurements of the glycerol solution (Fig.~\ref{fig:glycerol_20C}) due to exceeding the available torque range of the Brookfield viscometer. Within the available torque range of the Brookfield viscometer (10--90\%), the discrepancy with the Anton Paar rheometer data averaged 4.4\% for the UL Adapter and about 4\% for the LV3 spindle Table~\ref{tab:kcorr_all}. Thus, the correction coefficients closest to unity were obtained for the LV3 spindle and the UL Adapter measurement system (Fig.~\ref{fig:kcorr_glycerol}). The average values of the corrected coefficients obtained are presented in Table~\ref{tab:kcorr_all}. The Reynolds number did not exceed unity over the entire shear rate range ($Re$$<$1), indicating that the laminar flow conditions were maintained.}

{It is also noted in the literature that when the viscosity of highly concentrated glycerol solutions ($>$90\%) is measured using rotational rheometers, unstable viscosity values are observed at low shear rates ($\dot{\gamma}$ $<$ 1 \si{s^{-1}}), in contrast to the stable viscosity values obtained at shear rates $\dot{\gamma}$ $>$ 1 \si{s^{-1}}. This phenomenon is interpreted as a limitation of the instrument measurement sensitivity and the influence of unsteady flow effects rather than as the true non-Newtonian behavior of pure glycerol \citep{bib1}. An increase in glycerol viscosity was also observed at shear rates $\dot{\gamma}$ $<$ 0.1 \si{s^{-1}} using the Anton Paar MCR 302 rheometer with the CC27 measuring system; however, this result was reserved for further detailed investigation, as the Brookfield viscometer was no longer applicable in this shear rate range and therefore refinement of its measurement results was not required.}

{Therefore, using the specified configuration, the shear rate ranges from 0.22 to 103.5 \si{s^{-1}} can be covered for a glycerol solution with a concentration of 92.5\%.}

\begin{figure}[H]
\centering
\includegraphics[width=\textwidth]{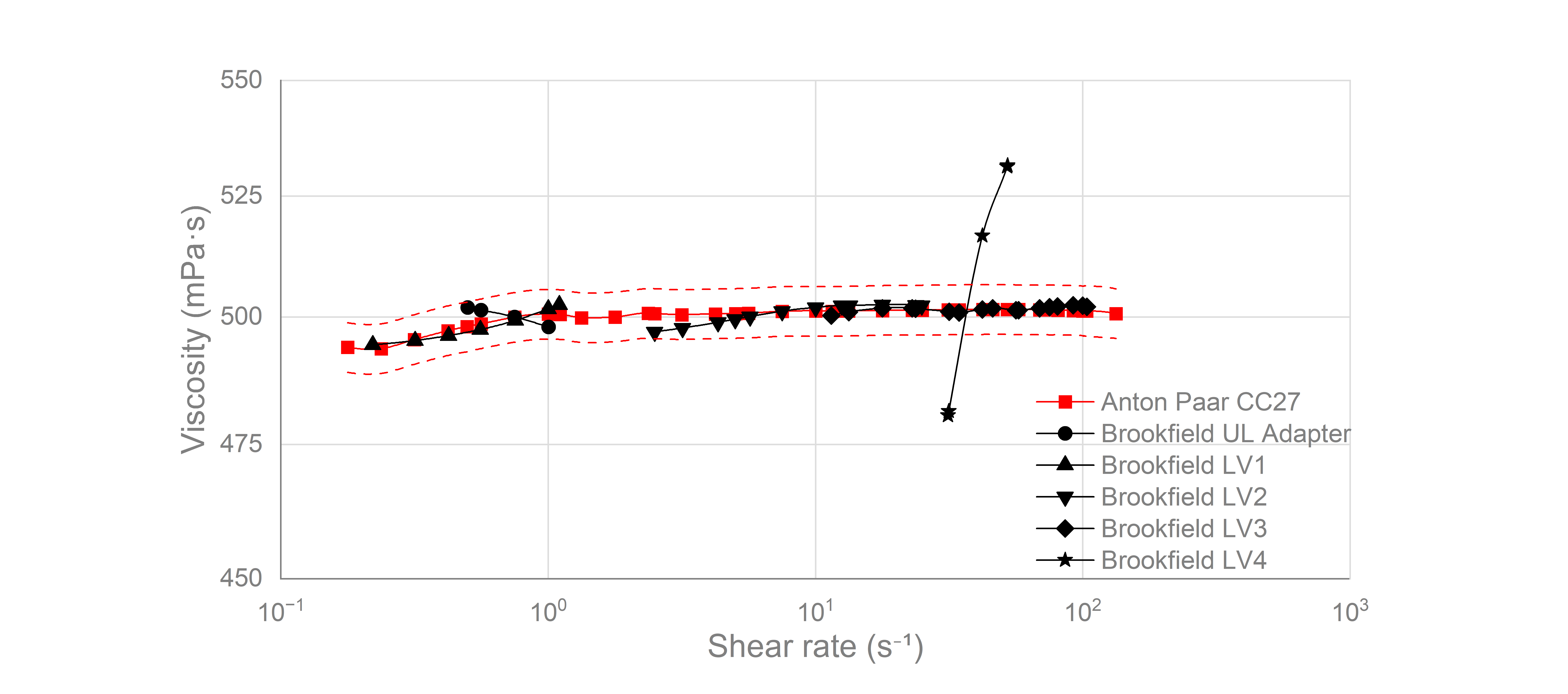}
\caption{Viscosity as a function of shear rate for a 92.5\% glycerol solution at 20~\si{\degreeCelsius} after applying the correction coefficient $k_{\mathrm{corr}}$ to the Brookfield viscometer data. The red dashed lines show the confidence interval ±1\%.}
\label{fig:glycerol_corrected}
\end{figure}

{Application of the correction coefficient provides good quantitative agreement between viscosity measurements performed using the UL Adapter and LV1–LV3 systems of the Brookfield viscometer and the reference values obtained with the Anton Paar MCR 302 rheometer equipped with the CC27 measuring system, within a ±1\% confidence interval throughout the investigated shear rate range (Fig.~\ref{fig:glycerol_corrected}). This result indicates the validity of converting the Brookfield viscometer readings using the coefficients obtained and confirms the possibility of their use for quantitative analysis under the specified flow conditions. For the LV4 spindle, systematic deviations beyond the confidence interval were observed, indicating that the measurements fall outside the range of conditions under which the proposed correction coefficient remains applicable. Thus, the use of the LV4 spindle requires additional caution when investigating Newtonian fluids with viscosities insufficiently high for this spindle and cannot be regarded as equivalent to other measurement systems without additional analysis of the flow regimes.}

\subsubsection{Results of Testing a 50\% Glycerol Solution}\label{subsubsec2}

\begin{figure}[H]
\centering
\includegraphics[width=\textwidth]{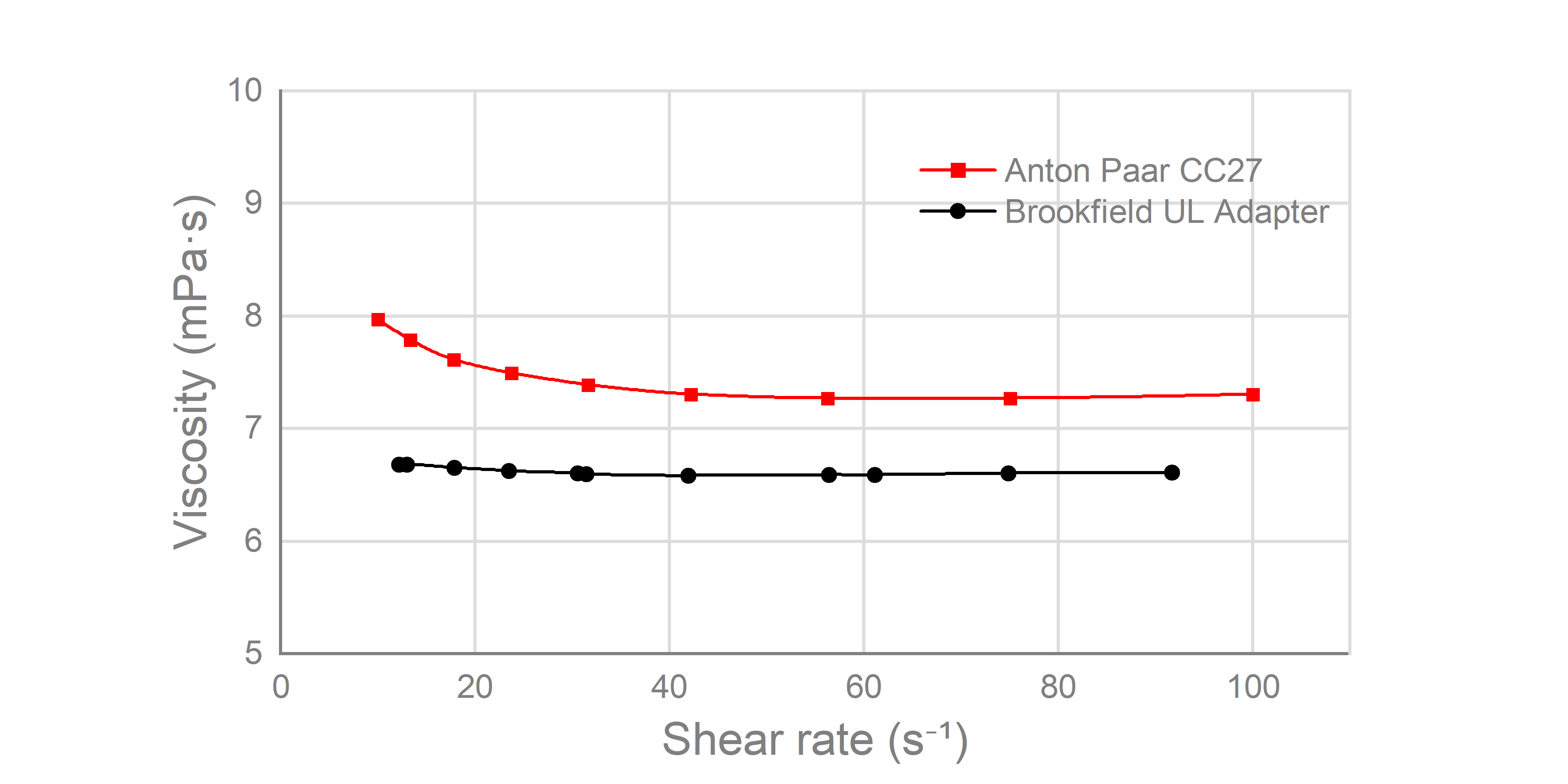}
\caption{Viscosity as a function of shear rate for a 50\% glycerol solution at 20~\si{\degreeCelsius}, measured using the Anton Paar rheometer (CC27) and the Brookfield viscometer with the UL Adapter.}
\label{fig:glycerol50}
\end{figure}

\begin{figure}[H]
\centering
\includegraphics[width=\textwidth]{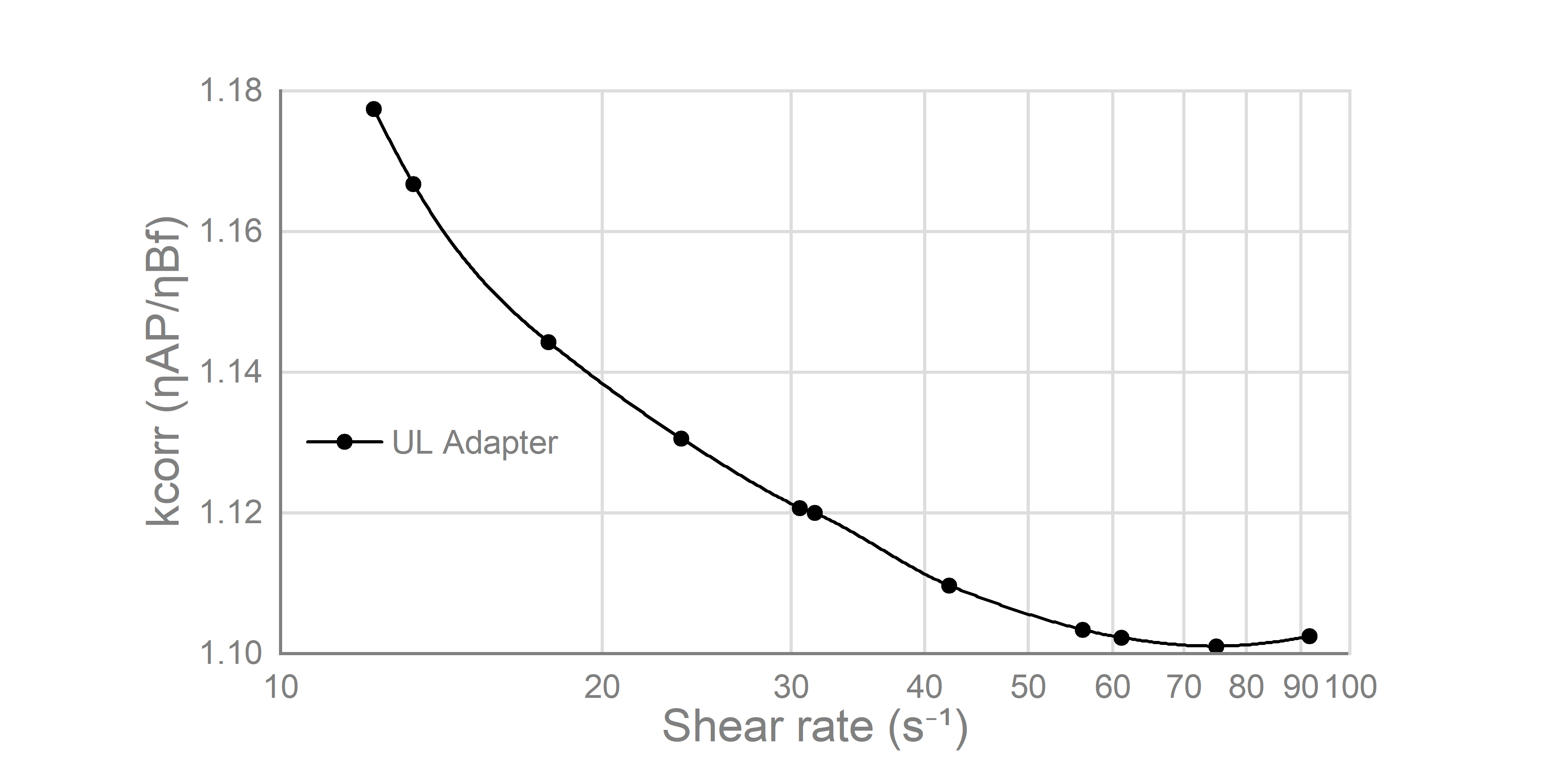}
\caption{Correction coefficient $k_{\mathrm{corr}} = \eta_{\mathrm{AP}} / \eta_{\mathrm{Bf}}$ as a function of shear rate for a 50\% glycerol solution at 20~\si{\degreeCelsius} obtained for different Brookfield spindle geometries.}
\label{fig:glycerol50_corr}
\end{figure}

{For the 50\% glycerol solution, Brookfield measurements were performed only using the UL Adapter system, since according to technical documentation the LV1–LV4 geometries are capable of measuring viscosity only when its expected value exceeds 15 \si{\milli\pascal\second} \citep{bib2,bib3} (Fig.~\ref{fig:glycerol50}). As in the case of the 92.5\% glycerol solution the measurements of the 50\% solution in \SI{20}{\degreeCelsius} showed discrepancies of up to 12.5\% compared to the reference data Table~\ref{tab:kcorr_all}, while the correction coefficient remained close to unity, as observed for the 92.5\% glycerol solution (Fig.~\ref{fig:glycerol50_corr}). The Reynolds number ranged from 3 to 22 and increased with shear rate, indicating a transitional flow regime and the occurrence of “end effects” \citep{bib7}. For the Anton Paar rheometer, $Re$ $>$ 1 was observed at shear rates $\dot{\gamma}$ $>$ 7.5 \si{s^{-1}}, indicating the presence of a transitional flow regime as mentioned by \cite{bib7}.}

{Thus, using the specified setup, the shear rate range of 12 to 92 \si{s^{-1}} can be covered for a solution of 50\% glycerol.}

\subsubsection{Distilled Water}\label{subsubsec2}

\begin{figure}[H]
\centering
\includegraphics[width=\textwidth]{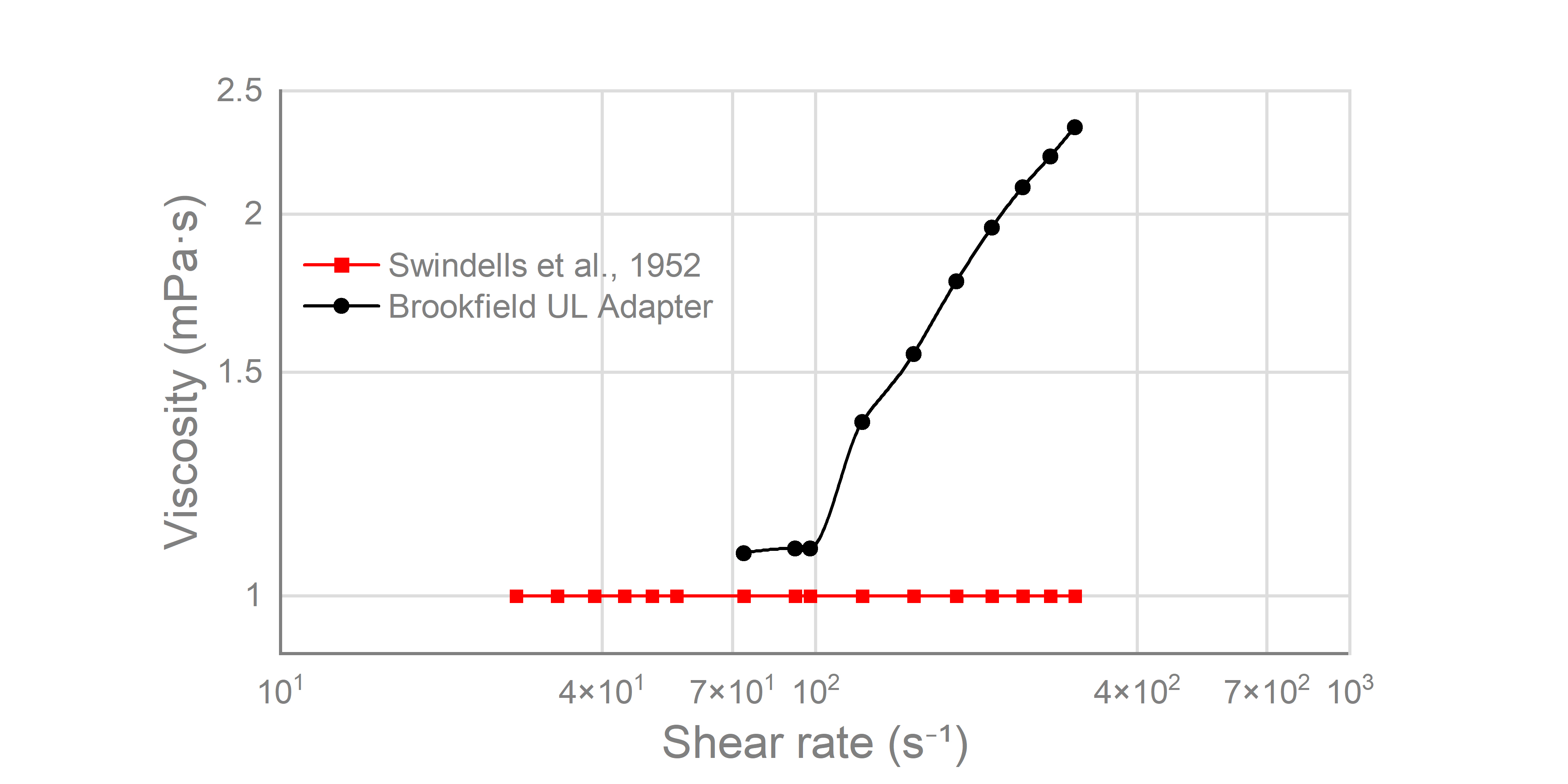}
\caption{Viscosity as a function of shear rate for distilled water at 20~\si{\degreeCelsius}, measured using the Brookfield viscometer with the UL Adapter and compared with reference data from \cite{bib14}.}
\label{fig:water}
\end{figure}

\begin{figure}[H]
\centering
\includegraphics[width=\textwidth]{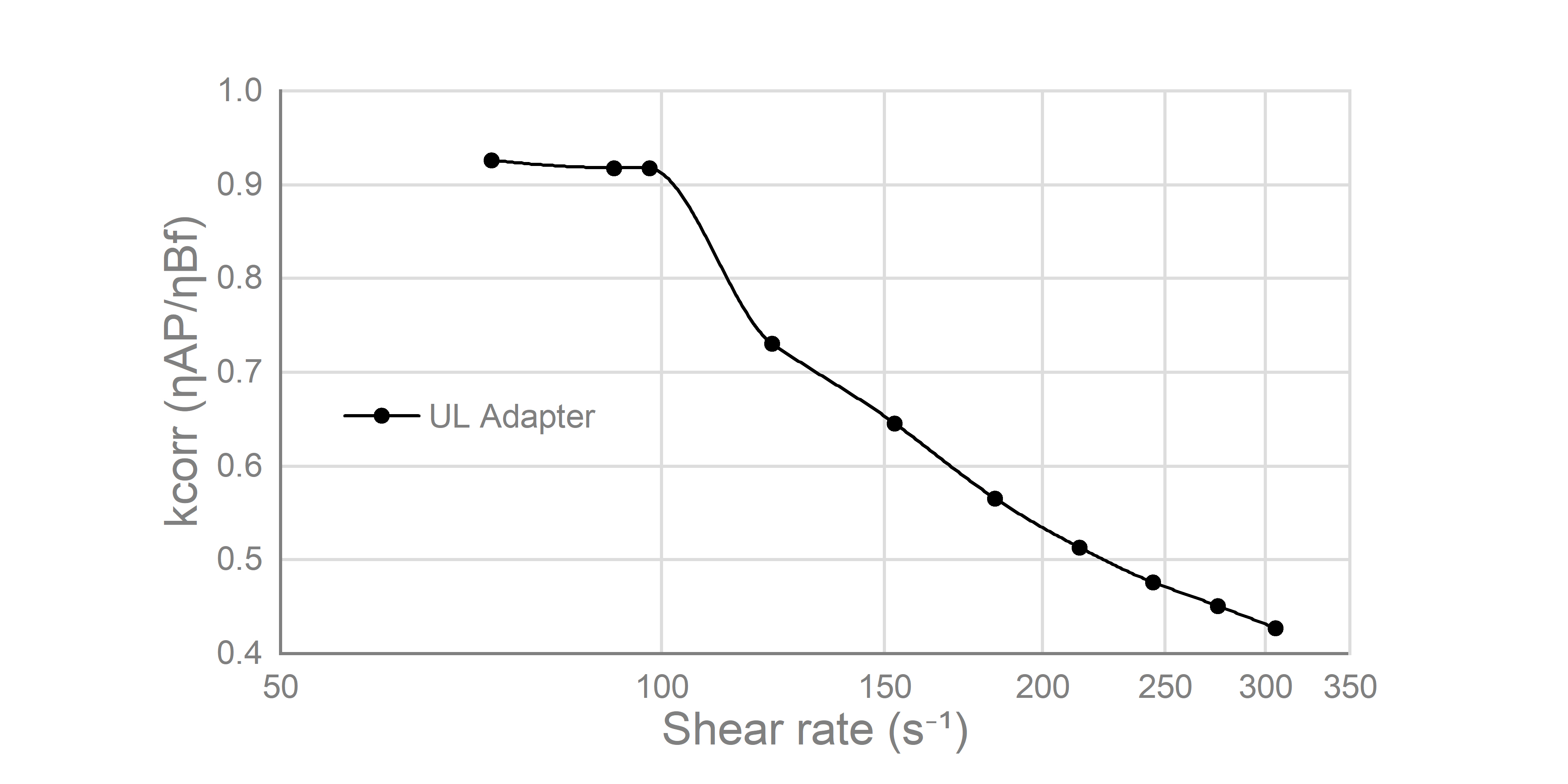}
\caption{Correction coefficient $k_{\mathrm{corr}} = \eta_{\mathrm{AP}} / \eta_{\mathrm{Bf}}$ as a function of shear rate for distilled water at 20~\si{\degreeCelsius} obtained using the UL Adapter.}
\label{fig:kcorr_water}
\end{figure}

{The viscosity of water at \SI{20}{\degreeCelsius} is approximately 1 \si{\milli\pascal\second} \citep{bib14}. According to technical documentation, the UL Adapter is the only measurement system capable of measuring viscosities from 1 \si{\milli\pascal\second} \citep{bib2,bib3}. In this regard, the behavior of the geometry at the lower boundary of the permissible viscosity range was investigated (Fig.~\ref{fig:water}). According to the results obtained, reliable values are achieved only at shear rates $\dot{\gamma}$ $\leq$ 100 \si{s^{-1}}. When this threshold is exceeded, an increase in the measured viscosity is observed, reaching up to 2.34 \si{\milli\pascal\second} at $\dot{\gamma}$ $\approx$ 306 \si{s^{-1}}. The minimum discrepancy of 7.4\% was obtained at $\dot{\gamma}$ = 73.38 \si{s^{-1}}, whereas the maximum exceeded 57\% at $\dot{\gamma}$ $\approx$ 306 \si{s^{-1}} (Fig.~\ref{fig:kcorr_water}). In this case, the average discrepancy and the correction coefficient were calculated only for the first three data points Table~\ref{tab:kcorr_all}. The Reynolds number ranged from 95 to 183 (95 $\leq$ $Re$ $\leq$ 183), indicating a transitional flow regime.}

{Hence, the use of the UL Adapter measurement system for distilled water and fluids with similar viscosity allows reliable values to be obtained within a relatively narrow shear rate range of 73 to 98 \si{s^{-1}}.}

\subsection{Viscosity Measurements of Non-Newtonian Fluids}\label{subsec2}

{As non-Newtonian fluids, guar gum powder–based solutions were investigated; the powder was mixed at room temperature using a mixer at 1500 rpm for 10 min. Subsequently, the solution was allowed to stand for 30 min, after which cross-linking was performed. During the cross-linking process, the cross-linking agent BCF-9 was introduced into the guar gum solution by syringe in continuous mixing with the mixer blades rotating at 1500 rpm, after which the solution was allowed to stand for an additional 1 h prior to the start of measurements. Based on the evaluation of the Reynolds number, it was found that for guar gum–based fluids $Re$ remained below 2 for all measurements, indicating a laminar flow regime during the experiments.}

\subsubsection{Linear Guar Gum–Based Gel}\label{subsubsec2}

\begin{figure}[H]
\centering
\includegraphics[width=\textwidth]{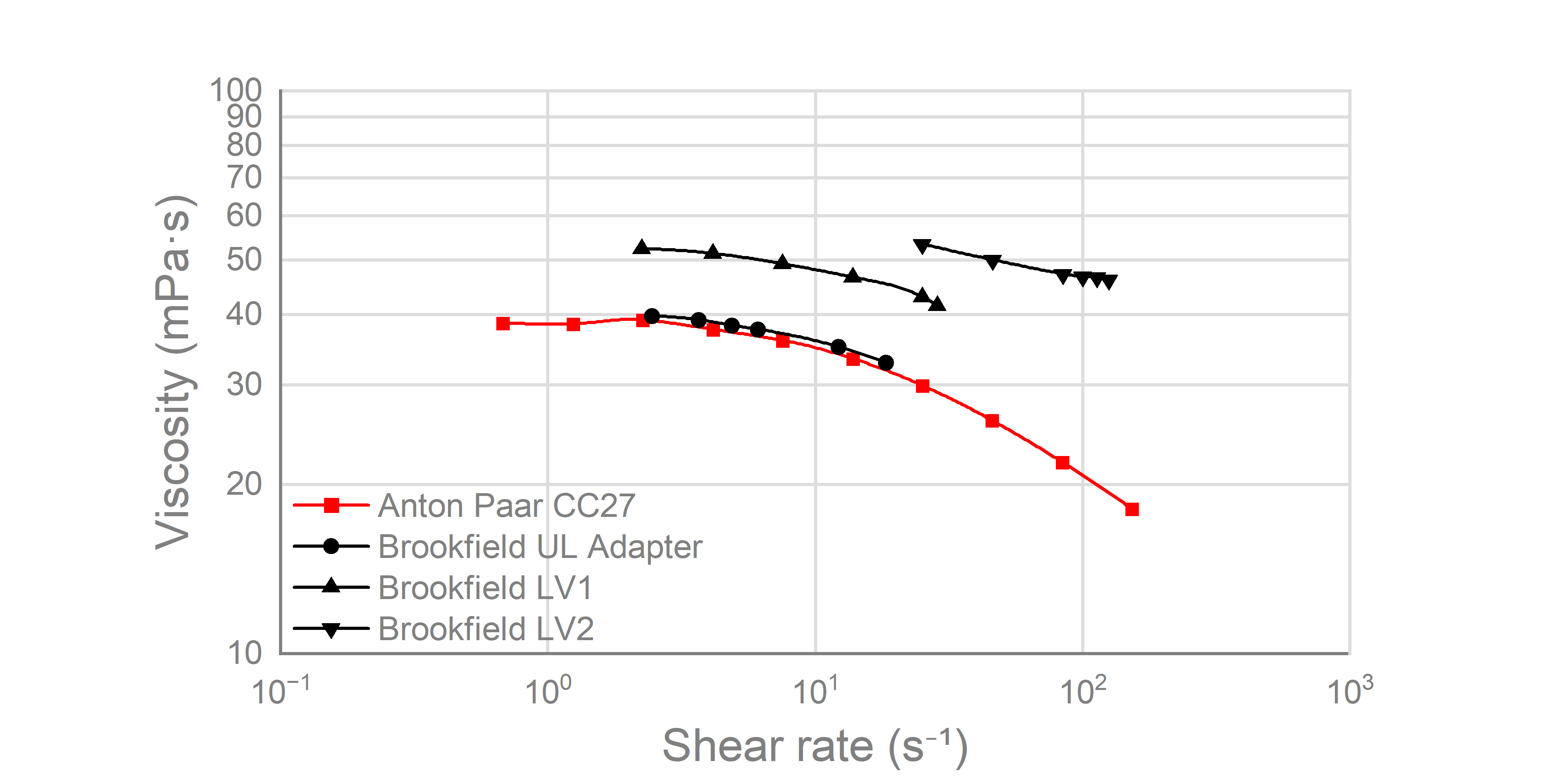}
\caption{Comparison of viscosity measurements for a linear guar gel (2.4~\si{\kilogram\per\meter\cubed}) at 20~\si{\degreeCelsius} obtained using Brookfield spindles (LV1, LV2, UL Adapter) and the Anton Paar CC27 measuring system.}
\label{fig:guar24_20}
\end{figure}

\begin{figure}[H]
\centering
\includegraphics[width=\textwidth]{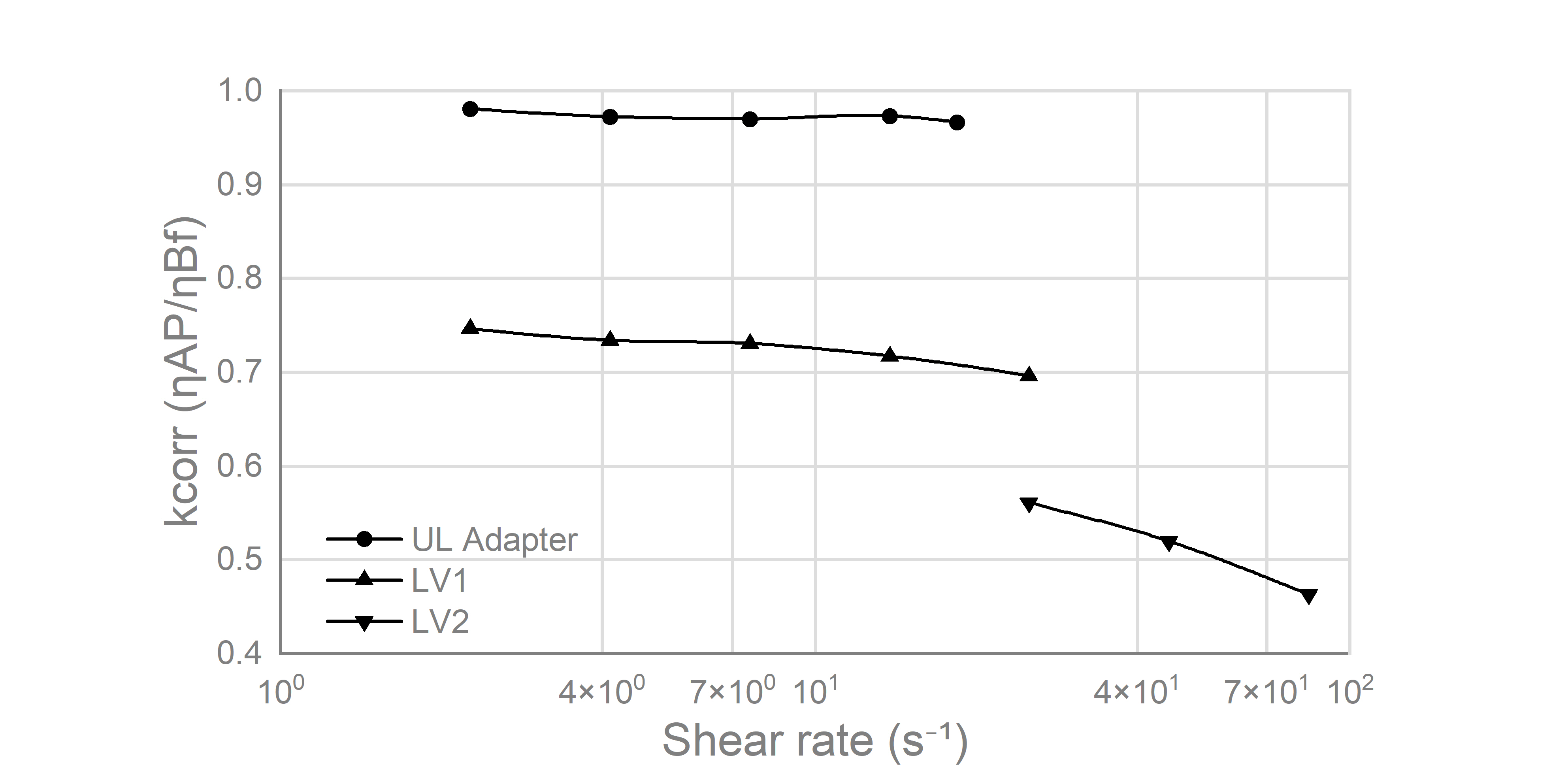}
\caption{Dependence of the correction coefficient $k_{\mathrm{corr}}$ on shear rate for a linear guar gel (2.4~\si{\kilogram\per\meter\cubed}) at 20~\si{\degreeCelsius} for different Brookfield measuring systems (UL Adapter, LV1, LV2) relative to Anton Paar CC27.}
\label{fig:guar24_20_kcorr}
\end{figure}

\begin{figure}[H]
\centering
\includegraphics[width=\textwidth]{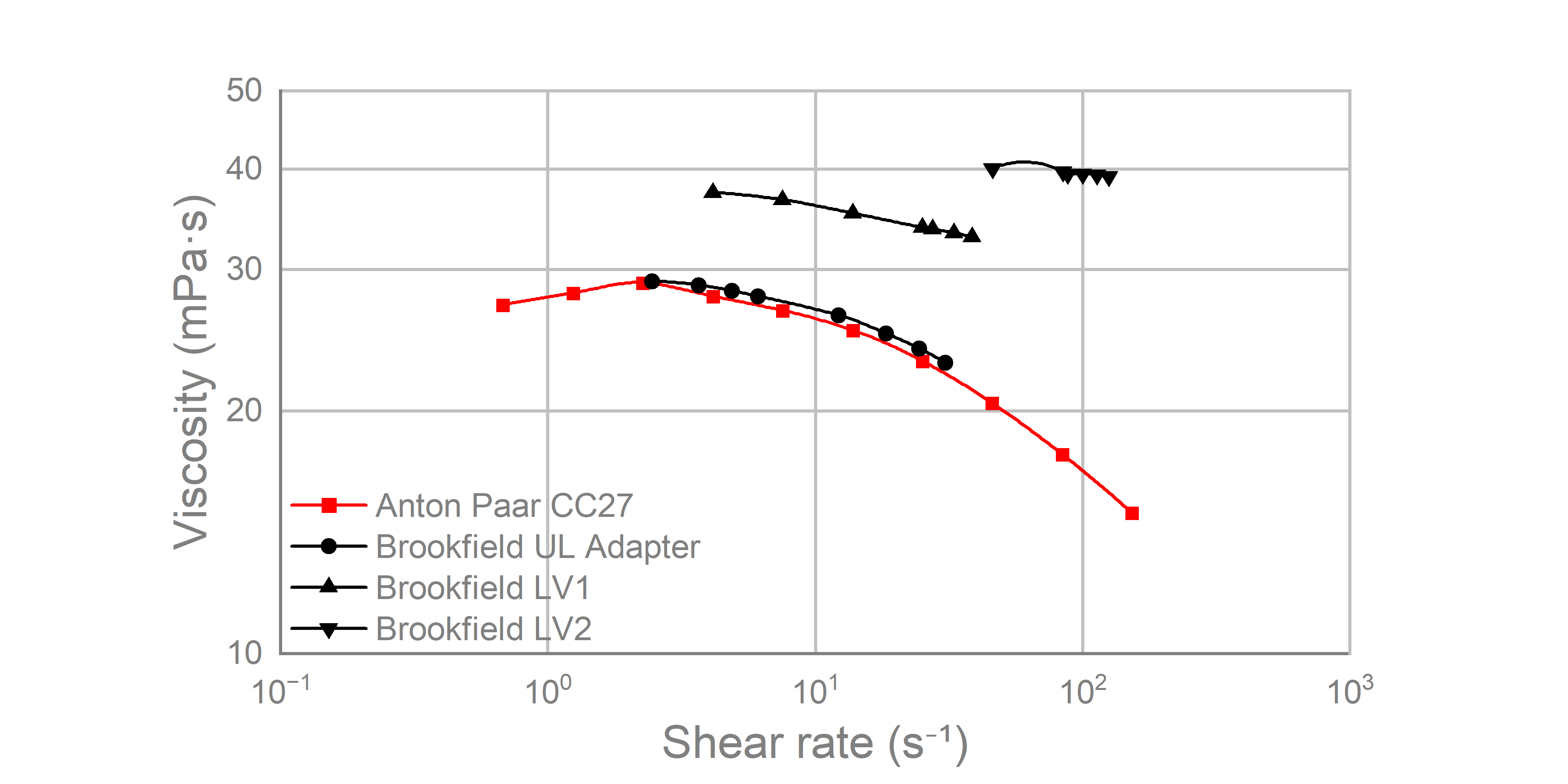}
\caption{Comparison of viscosity measurements for a linear guar gel (2.4~\si{\kilogram\per\meter\cubed}) at 30~\si{\degreeCelsius} obtained using Brookfield spindles (LV1, LV2, UL Adapter) and the Anton Paar CC27 measuring system.}
\label{fig:guar24_30}
\end{figure}

\begin{figure}[H]
\centering
\includegraphics[width=\textwidth]{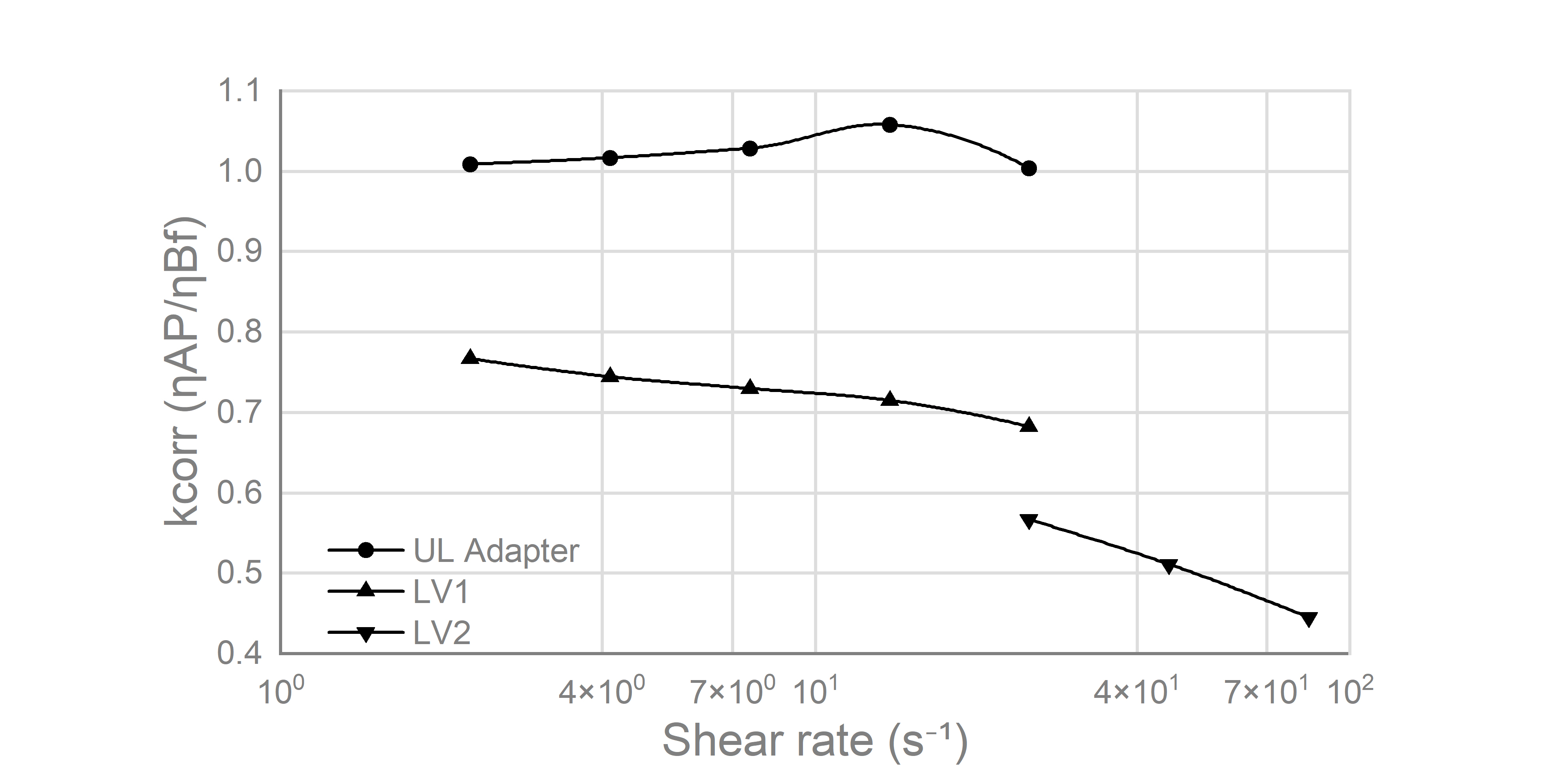}
\caption{Dependence of the correction coefficient $k_{\mathrm{corr}}$ on shear rate for a linear guar gel (2.4~\si{\kilogram\per\meter\cubed}) at 30~\si{\degreeCelsius} for different Brookfield measuring systems (UL Adapter, LV1, LV2) relative to Anton Paar CC27.}
\label{fig:guar24_30_kcorr}
\end{figure}

\begin{figure}[H]
\centering
\includegraphics[width=\textwidth]{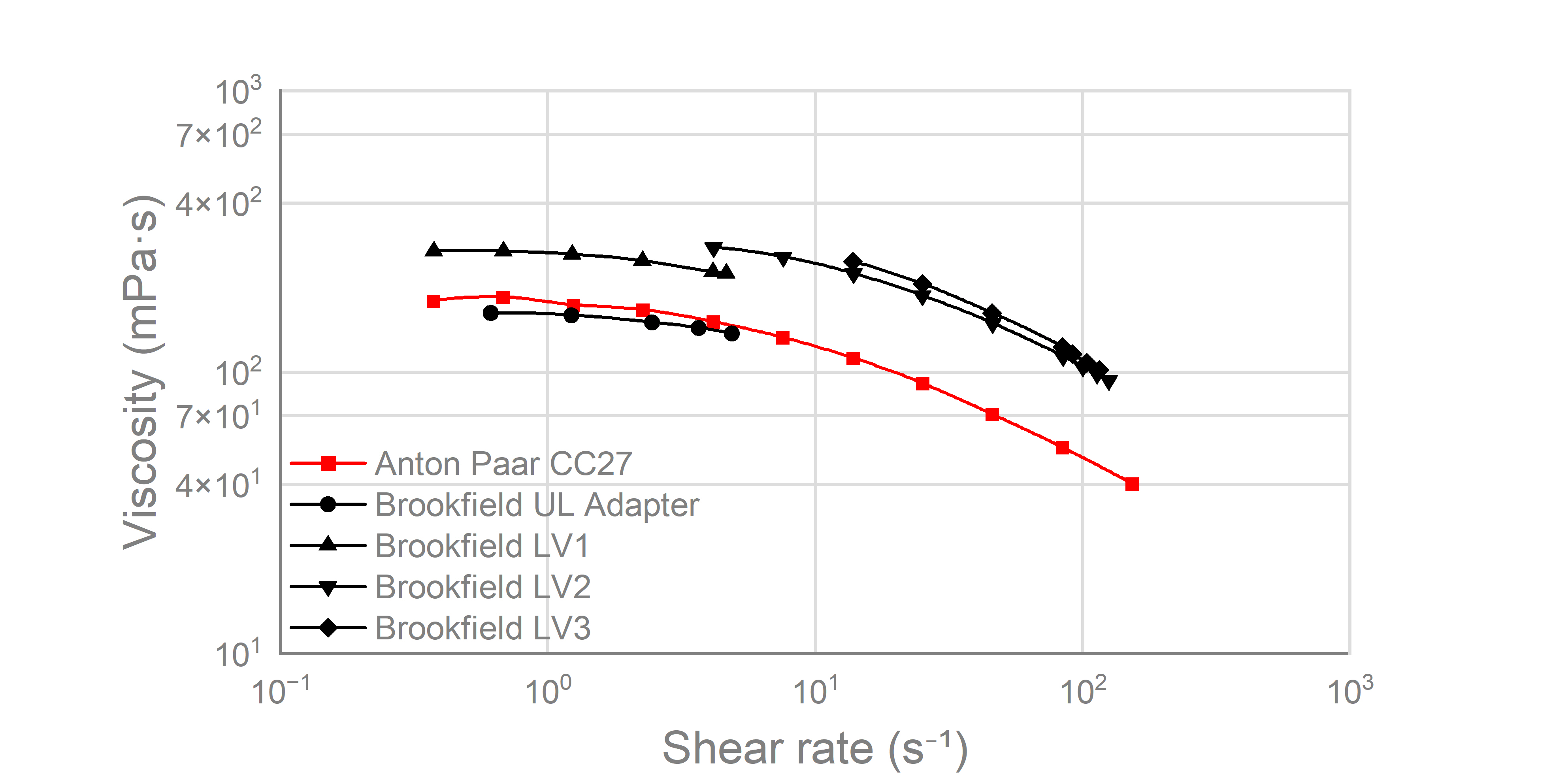}
\caption{Comparison of viscosity measurements for a linear guar gel (3.6~\si{\kilogram\per\meter\cubed}) at 20~\si{\degreeCelsius} obtained using Brookfield spindles (LV1, LV2, LV3, UL Adapter) and the Anton Paar CC27 measuring system.}
\label{fig:guar36_20}
\end{figure}

\begin{figure}[H]
\centering
\includegraphics[width=\textwidth]{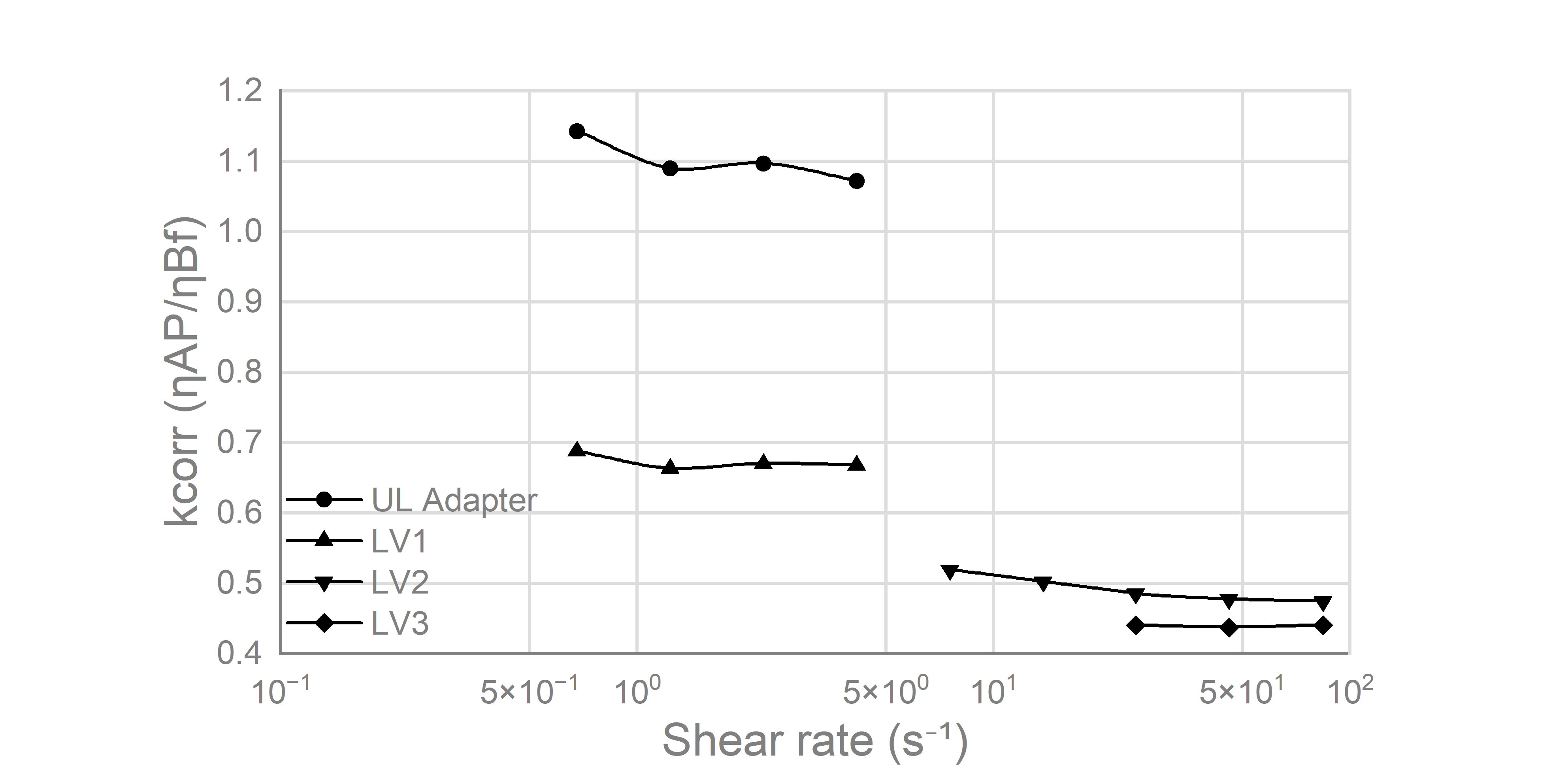}
\caption{Dependence of the correction coefficient $k_{\mathrm{corr}}$ on shear rate for a linear guar gel (3.6~\si{\kilogram\per\meter\cubed}) at 20~\si{\degreeCelsius} for different Brookfield measuring systems (UL Adapter, LV1, LV2, LV3) relative to Anton Paar CC27.}
\label{fig:guar36_20_kcorr}
\end{figure}

\begin{figure}[H]
\centering
\includegraphics[width=\textwidth]{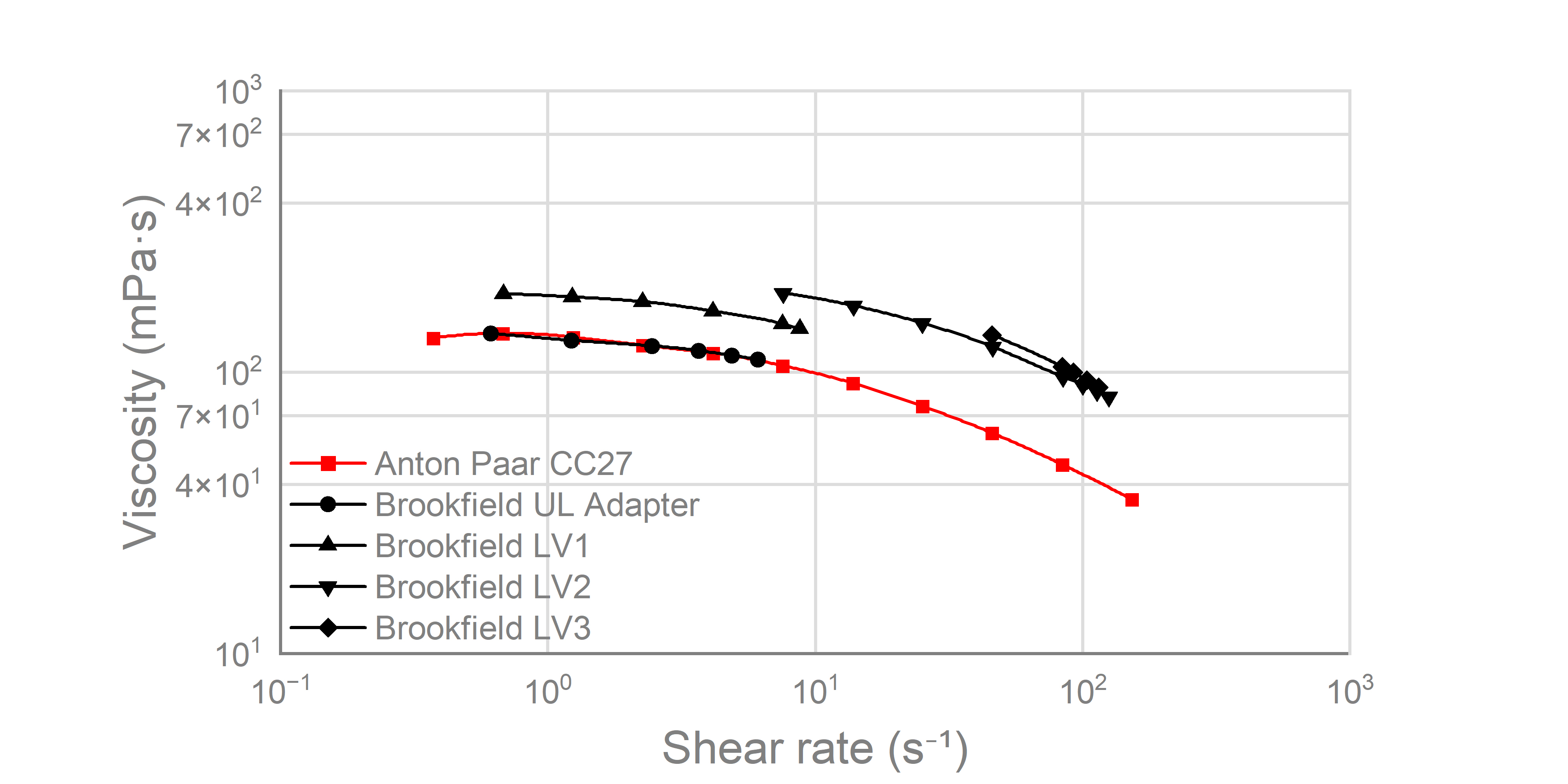}
\caption{Comparison of viscosity measurements for a linear guar gel (3.6~\si{\kilogram\per\meter\cubed}) at 30~\si{\degreeCelsius} obtained using Brookfield spindles (LV1, LV2, LV3, UL Adapter) and the Anton Paar CC27 measuring system.}
\label{fig:guar36_30}
\end{figure}

\begin{figure}[H]
\centering
\includegraphics[width=\textwidth]{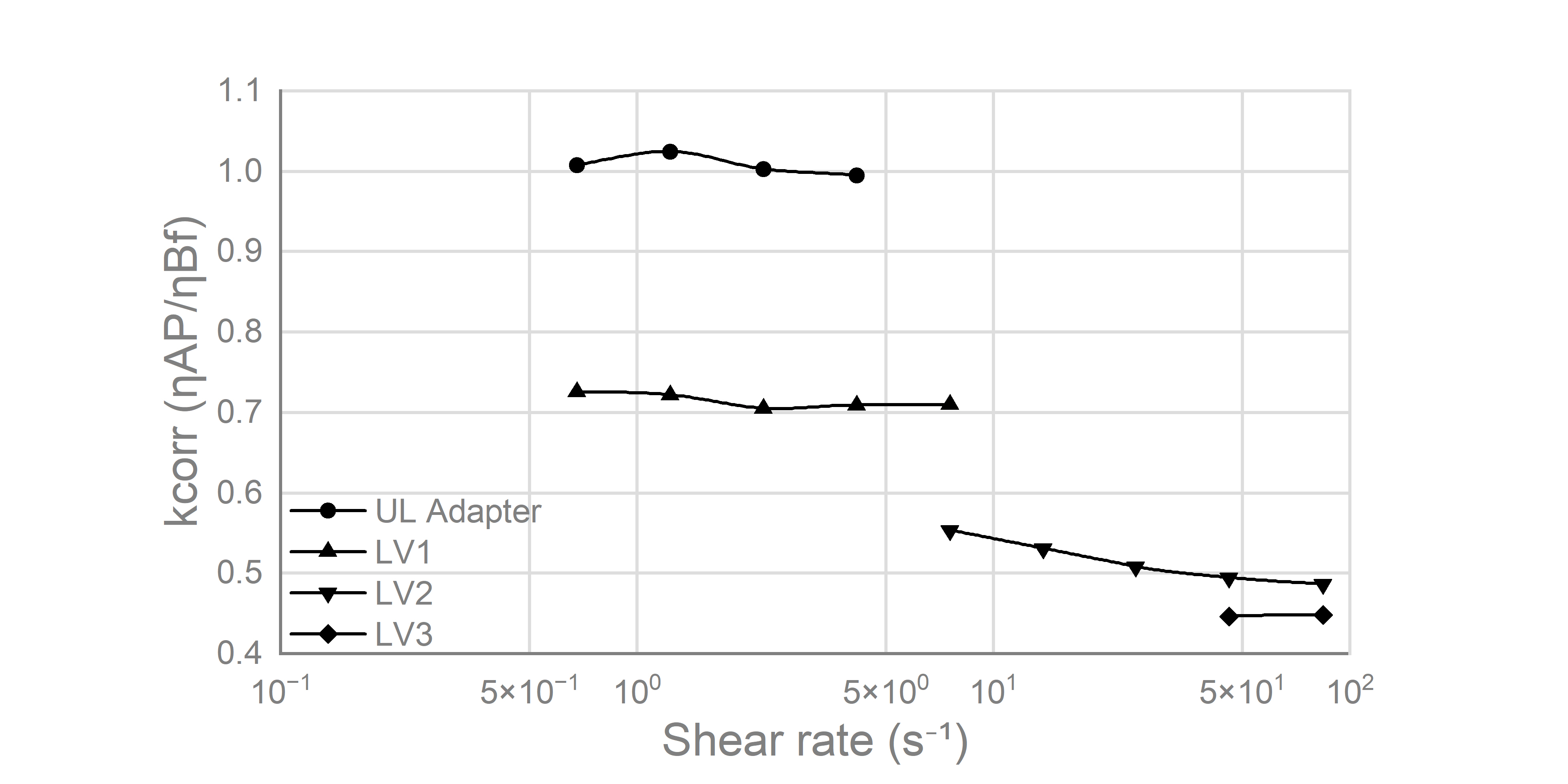}
\caption{Dependence of the correction coefficient $k_{\mathrm{corr}}$ on shear rate for a linear guar gel (3.6~\si{\kilogram\per\meter\cubed}) at 30~\si{\degreeCelsius} for different Brookfield measuring systems (UL Adapter, LV1, LV2, LV3) relative to Anton Paar CC27.}
\label{fig:guar36_30_kcorr}
\end{figure}

{During the testing of linear guar gels prepared from WGA NG-1 guar gum, solutions with concentrations of 2.4 and 3.6 \si{\kilogram\per\meter\cubed} were used. Measurements were performed at temperatures of 20 and 30~\si{\degreeCelsius}. During the tests, the LV1–LV3 setups and the UL Adapter were used wherever applicable. The use of the LV4 spindle proved impossible due to its low sensitivity. The best results were obtained using the UL Adapter system, as evidenced by the plots presented (Figs.~\ref{fig:guar24_20}--\ref{fig:guar36_30}) and the table of coefficients Table~\ref{tab:kcorr_all}. When measuring the viscosity of linear guar gels, this measurement system is recommended as it is the most accurate and sensitive. In addition, the correction coefficient $k_{corr}$ remains close to unity in the case of the linear gel as well (Figs.~\ref{fig:guar24_20_kcorr}--\ref{fig:guar36_30_kcorr}).}

{Using the specified setups, a relatively wide shear rate range from 1.22 to 126 \si{s^{-1}} can be covered for a guar gum gel with a concentration of 2.4 \si{\kilogram\per\meter\cubed}. For a guar gum–based gel with a concentration of 3.6 \si{\kilogram\per\meter\cubed}, the specified set of geometries allows the shear rate range from 0.37 to 126 \si{s^{-1}} at 20~\si{\degreeCelsius} and from 0.61 to 126 \si{s^{-1}} at 30~\si{\degreeCelsius} to be covered.}

\subsubsection{Cross-linked Guar Gum–Based Gel}\label{subsubsec2}

\begin{figure}[H]
\centering
\includegraphics[width=\textwidth]{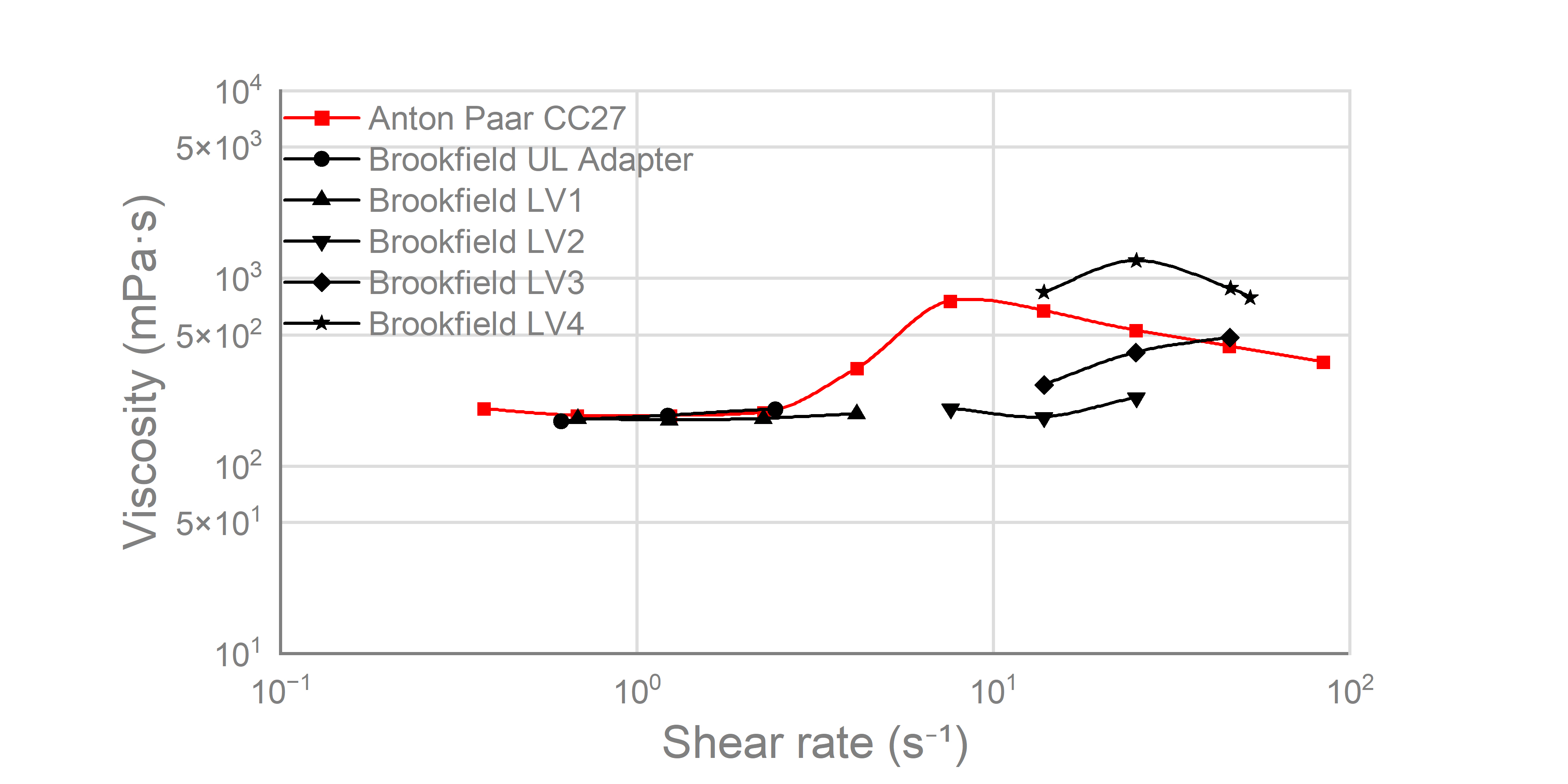}
\caption{Comparison of viscosity measurements for a cross-linked guar gel (2.4~\si{\kilogram\per\meter\cubed}) at 70~\si{\degreeCelsius} obtained using Brookfield spindles (LV1–LV4, UL Adapter) and the Anton Paar CC27 measuring system.}
\label{fig:guar_cross_70}
\end{figure}

\begin{figure}[H]
\centering
\includegraphics[width=\textwidth]{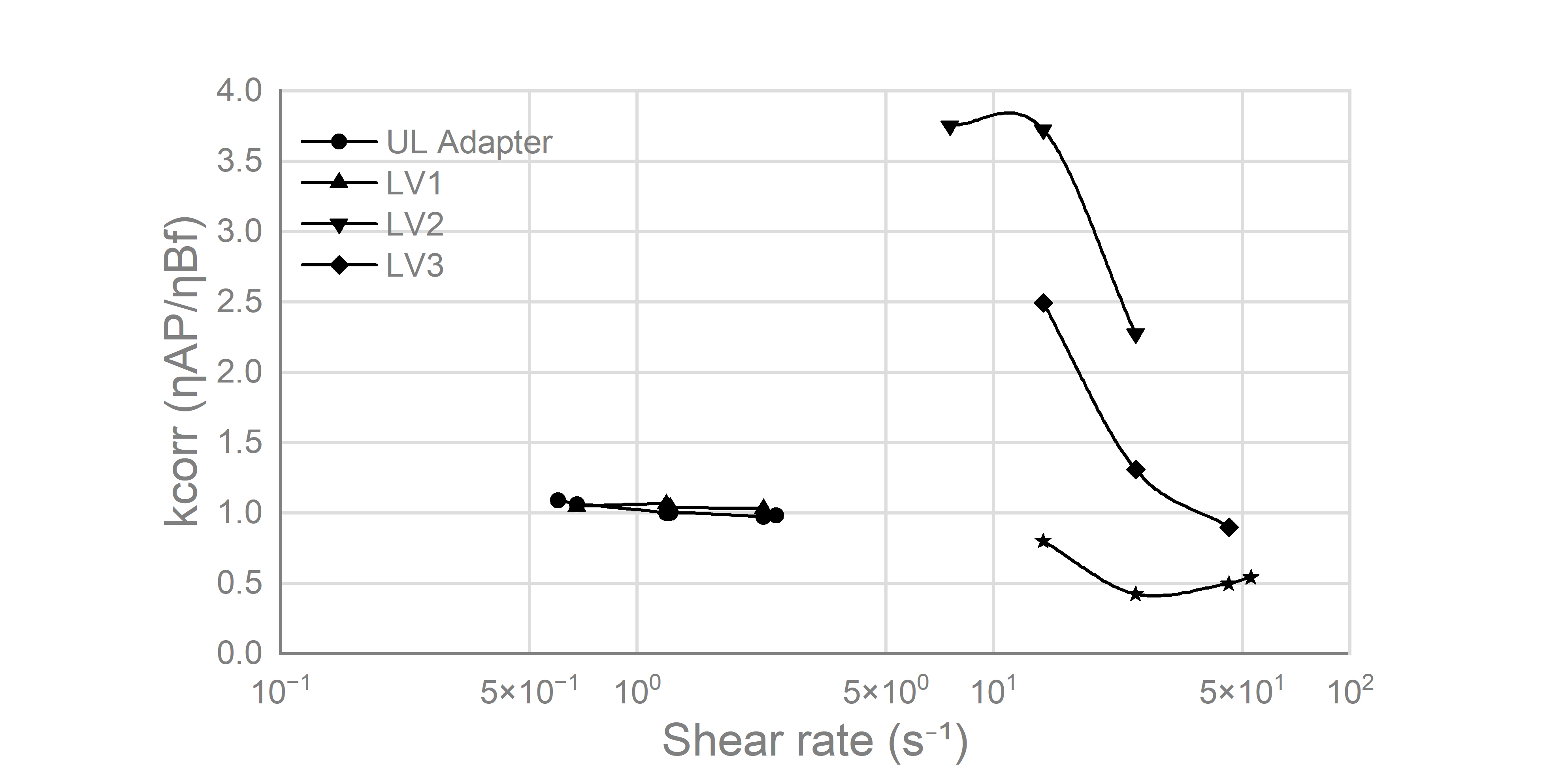}
\caption{Dependence of the correction coefficient $k_{\mathrm{corr}}$ on shear rate for a cross-linked guar gel (2.4~\si{\kilogram\per\meter\cubed}) at 70~\si{\degreeCelsius} for different Brookfield spindle geometries.}
\label{fig:kcorr_guar_cross_70}
\end{figure}

\begin{figure}[H]
\centering
\includegraphics[width=\textwidth]{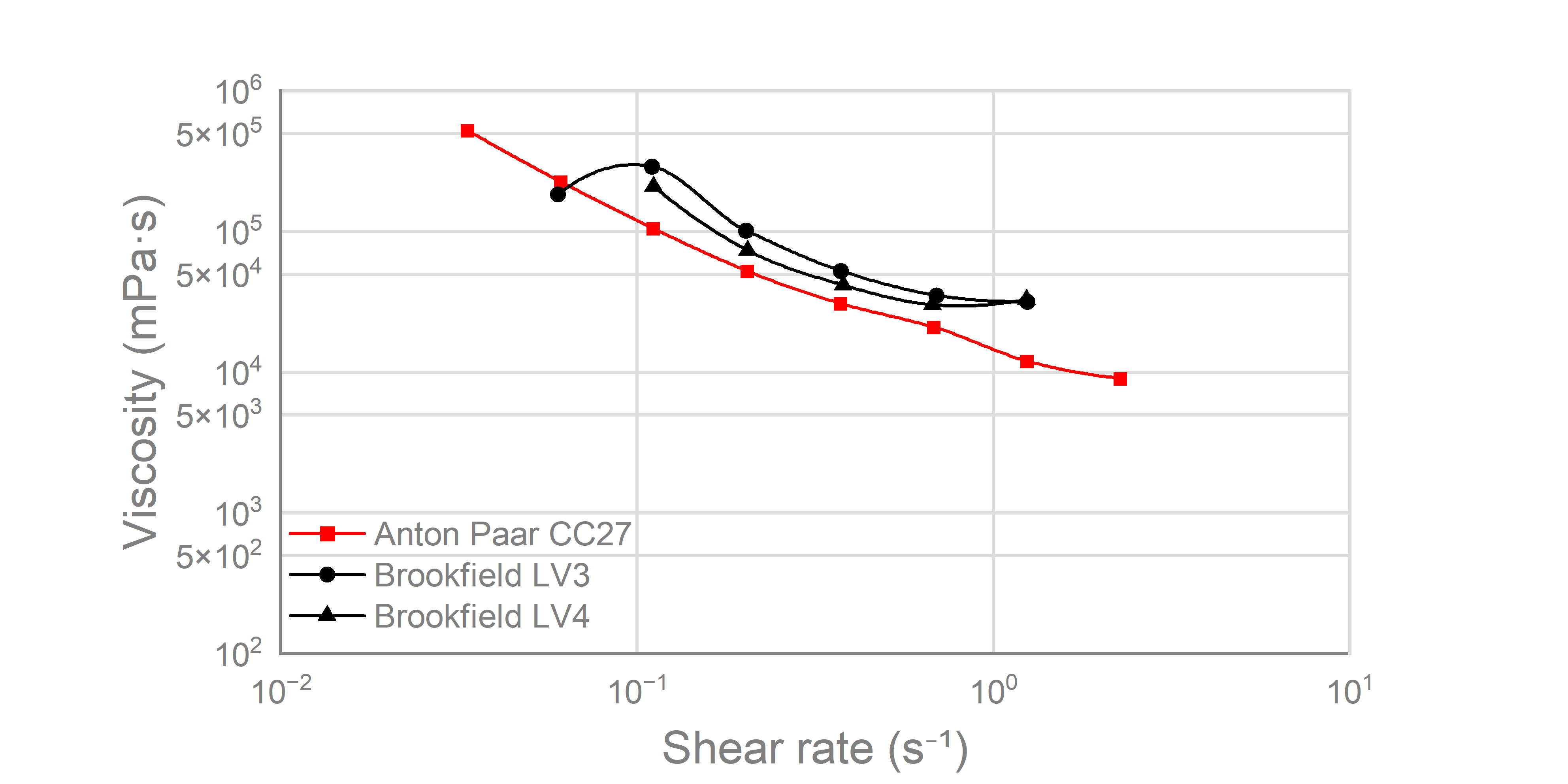}
\caption{Comparison of viscosity measurements for a cross-linked guar gel (2.4~\si{\kilogram\per\meter\cubed}) at 20~\si{\degreeCelsius} obtained using Brookfield spindles (LV3 and LV4) and the Anton Paar CC27 measuring system.}
\label{fig:guar_cross_20}
\end{figure}

\begin{figure}[H]
\centering
\includegraphics[width=\textwidth]{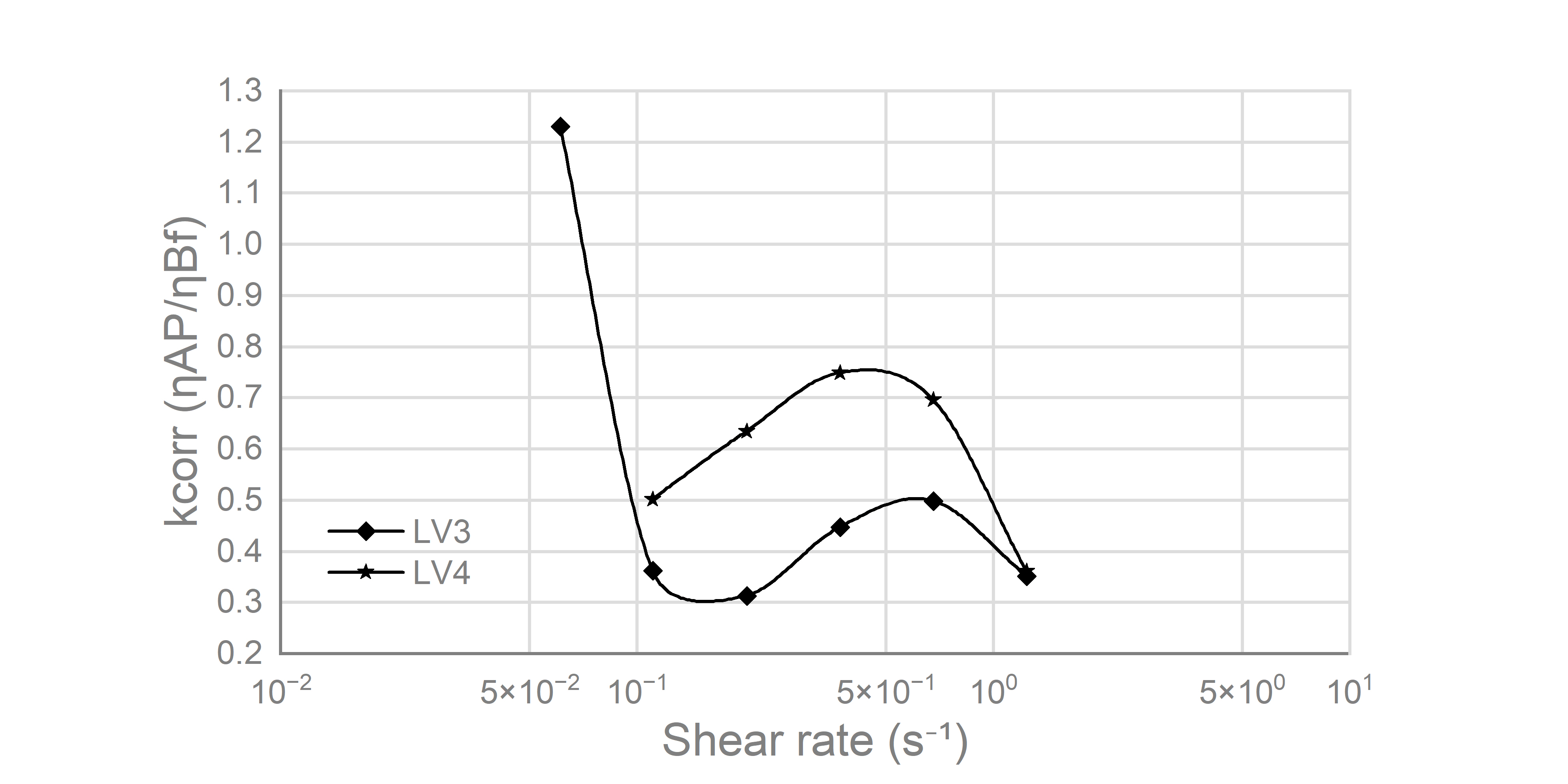}
\caption{Dependence of the correction coefficient $k_{\mathrm{corr}}$ on shear rate for a cross-linked guar gel (2.4~\si{\kilogram\per\meter\cubed}) at 20~\si{\degreeCelsius} for LV3 and LV4 spindles.}
\label{fig:kcorr_guar_cross_20}
\end{figure}

{Additionally, cross-linked gels based on WGA NG-1 guar gum and modified with the crosslinking agent BCF-9 were tested. Solutions with a gelling agent concentration of 2.4~\si{\kilogram\per\meter\cubed} were used, both under ambient conditions (T = 20~\si{\degreeCelsius}) and at the reservoir temperature (T = 70~\si{\degreeCelsius}) (Figs.~\ref{fig:guar_cross_70}, \ref{fig:guar_cross_20}). During the tests, the LV1–LV4 geometries and the UL Adapter were used. The use of the UL Adapter was only possible at 70~\si{\degreeCelsius} due to the high viscosity of the gel at T = 20~\si{\degreeCelsius}. The measurement results obtained using the LV3 measurement system at 20 and 70~\si{\degreeCelsius} differ significantly (Figs.~\ref{fig:kcorr_guar_cross_70}), \ref{fig:kcorr_guar_cross_20}. This observation indicates the need for caution when performing temperature-dependent measurements, particularly for cross-linked gels. This is also evidenced by the variation of the calculated correction coefficients for both temperatures Table~\ref{tab:kcorr_all}.}

{For guar gum gel with a concentration of 2.4~\si{\kilogram\per\meter\cubed}, the accessible shear rate range extends from 0.61 to 52.5 \si{s^{-1}} at 70~\si{\degreeCelsius} and from 0.06 to 1.24 \si{s^{-1}} at 20~\si{\degreeCelsius} when the specified configuration is used.}

\subsection{Applicability ranges of Brookfield measuring systems}\label{subsec2}

\begin{figure}[H]
\centering
\includegraphics[width=\textwidth]{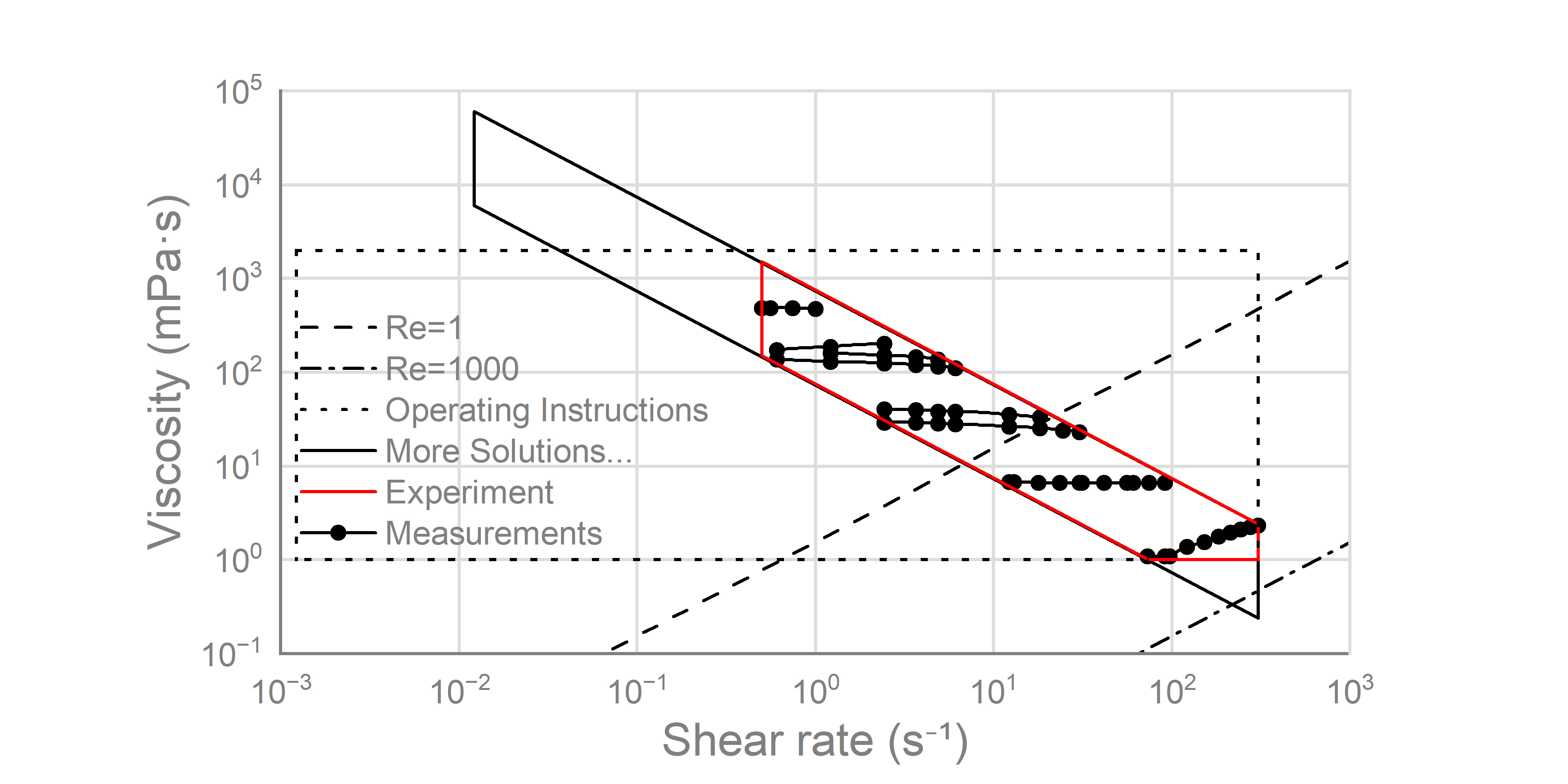}
\caption{Operating range of the Brookfield viscometer for the UL Adapter in viscosity–shear rate coordinates. The dashed lines correspond to Reynolds numbers $\mathrm{Re}=1$ and $\mathrm{Re}=1000$. The dotted rectangle indicates the manufacturer-recommended operating region. The solid lines represent theoretical viscosity limits. Experimental data are shown by symbols, and the red polygon highlights the range covered in the present study.}
\label{fig:ul_adapter_operating_range}
\end{figure}

\begin{figure}[H]
\centering
\includegraphics[width=\textwidth]{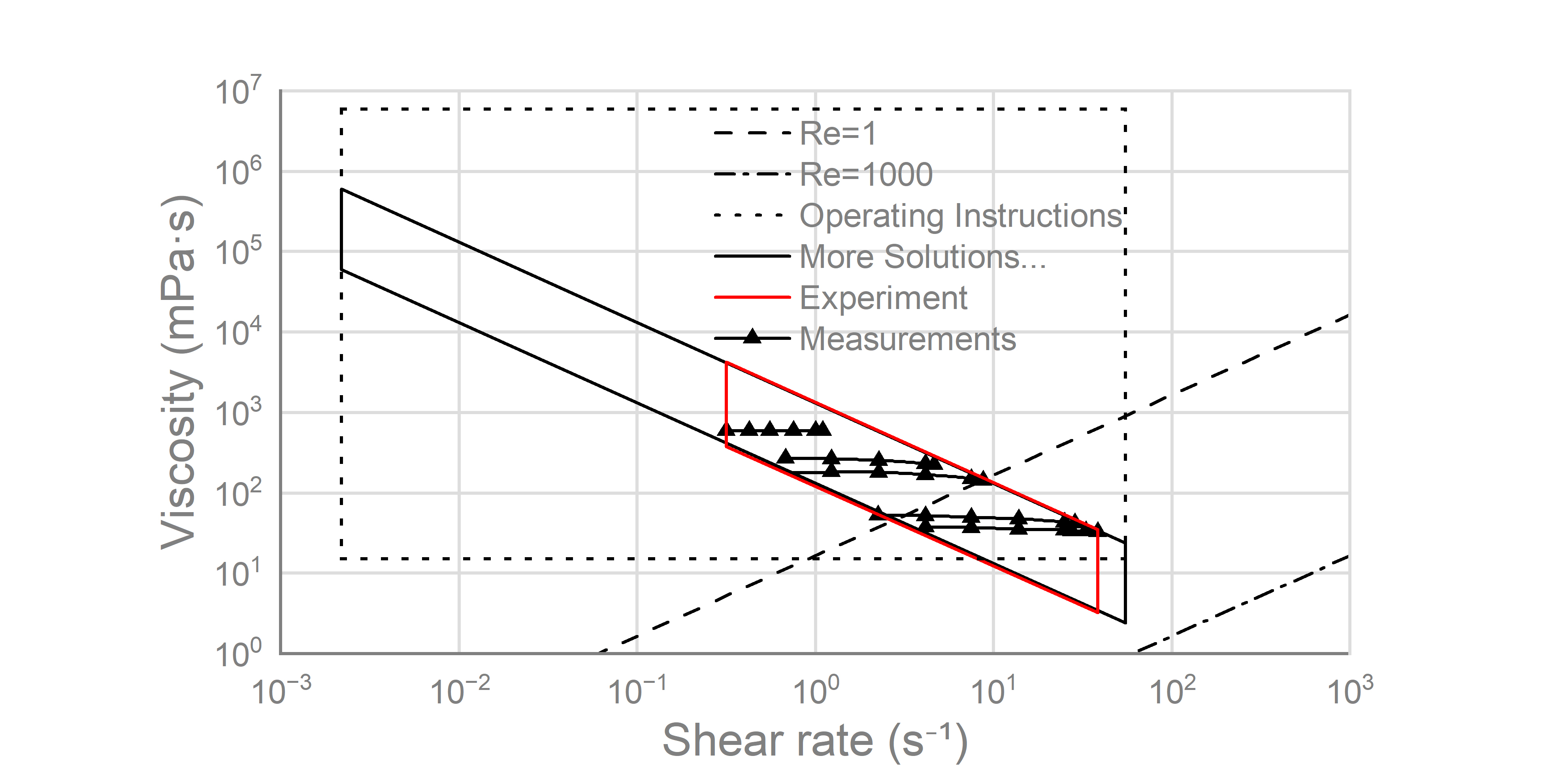}
\caption{Operating range of the Brookfield viscometer for the LV1 spindle in coordinates of viscosity versus shear rate. The dashed lines correspond to Reynolds numbers $\mathrm{Re}=1$ and $\mathrm{Re}=1000$. The dotted rectangle indicates the recommended operating region according to the manufacturer. The solid lines represent theoretical boundaries of measurable viscosity. Experimental data are shown by symbols, and the red polygon highlights the range covered in the present study.}
\label{fig:lv1_operating_range}
\end{figure}

\begin{figure}[H]
\centering
\includegraphics[width=\textwidth]{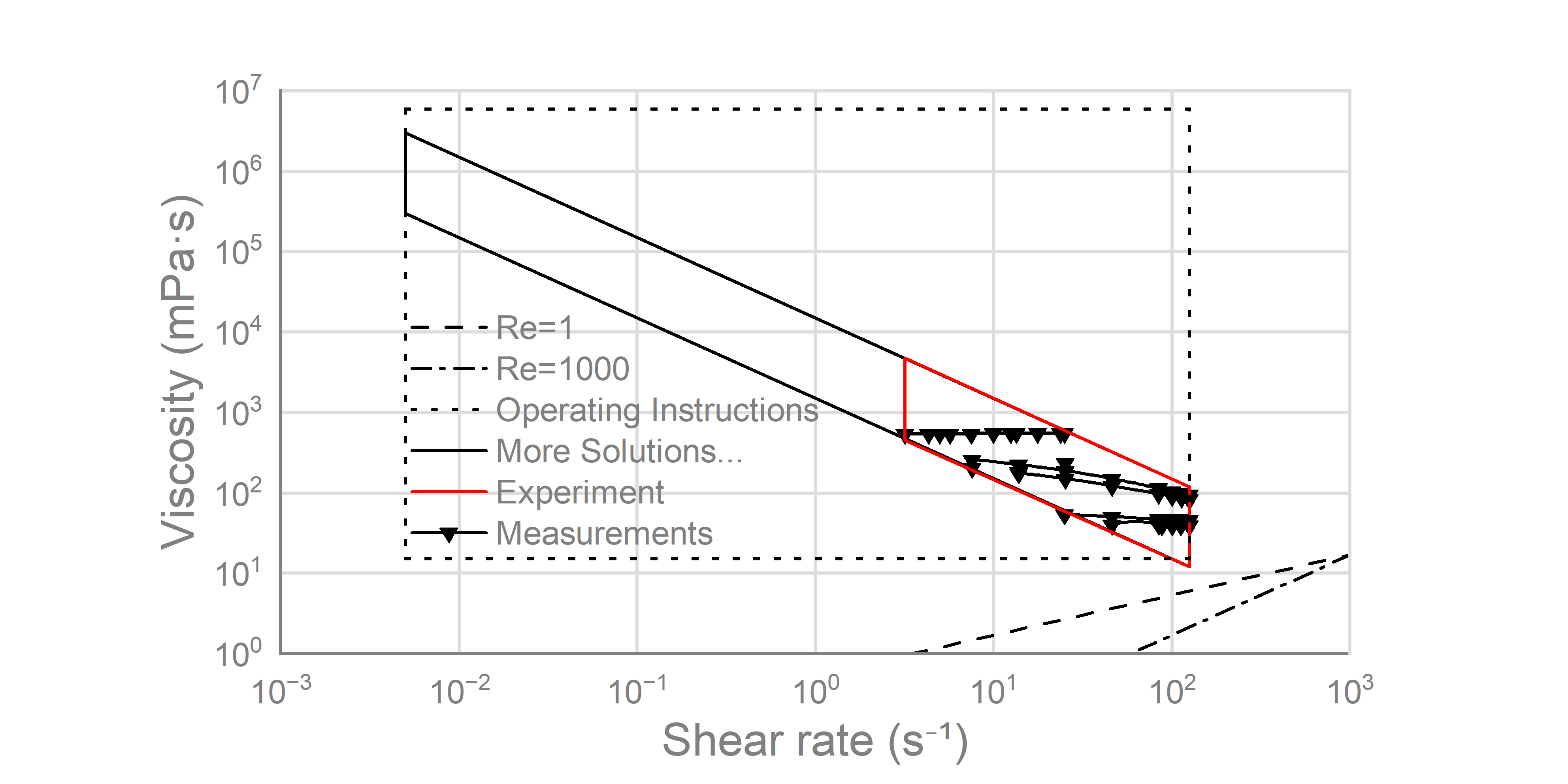}
\caption{Operating range of the Brookfield viscometer for the LV2 spindle in viscosity–shear rate coordinates. The dashed lines correspond to Reynolds numbers $\mathrm{Re}=1$ and $\mathrm{Re}=1000$. The dotted rectangle indicates the recommended operating region according to the manufacturer. The solid lines represent theoretical viscosity limits. Experimental data are shown by symbols, and the red polygon highlights the range covered in the present study.}
\label{fig:lv2_operating_range}
\end{figure}

\begin{figure}[H]
\centering
\includegraphics[width=\textwidth]{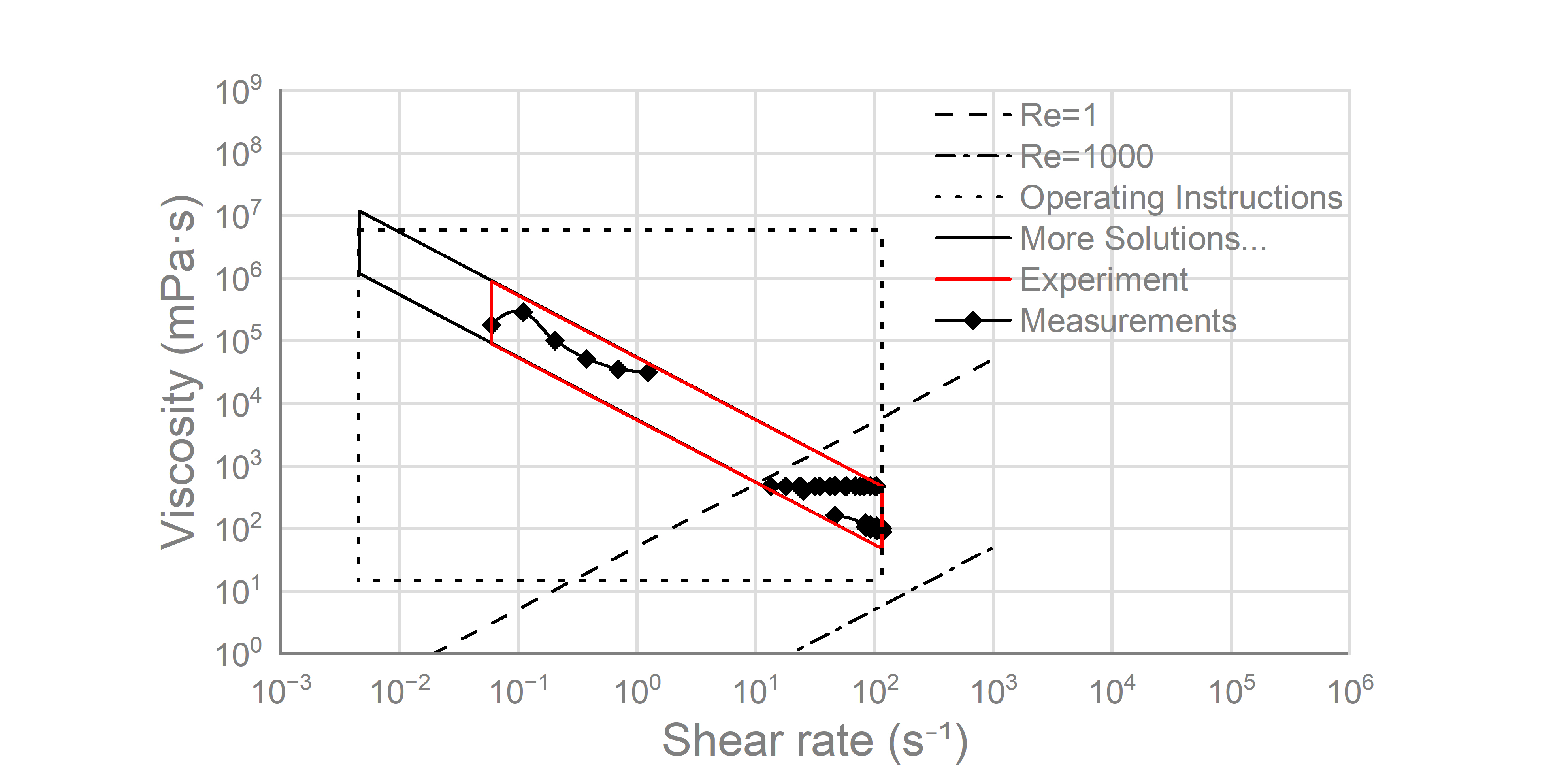}
\caption{Operating range of the Brookfield viscometer for the LV3 spindle in viscosity–shear rate coordinates. The dashed lines correspond to Reynolds numbers $\mathrm{Re}=1$ and $\mathrm{Re}=1000$. The dotted rectangle indicates the manufacturer-recommended operating region. The solid lines represent theoretical viscosity limits. Experimental data are shown by symbols, and the red polygon highlights the range covered in the present study.}
\label{fig:lv3_operating_range}
\end{figure}

\begin{figure}[H]
\centering
\includegraphics[width=\textwidth]{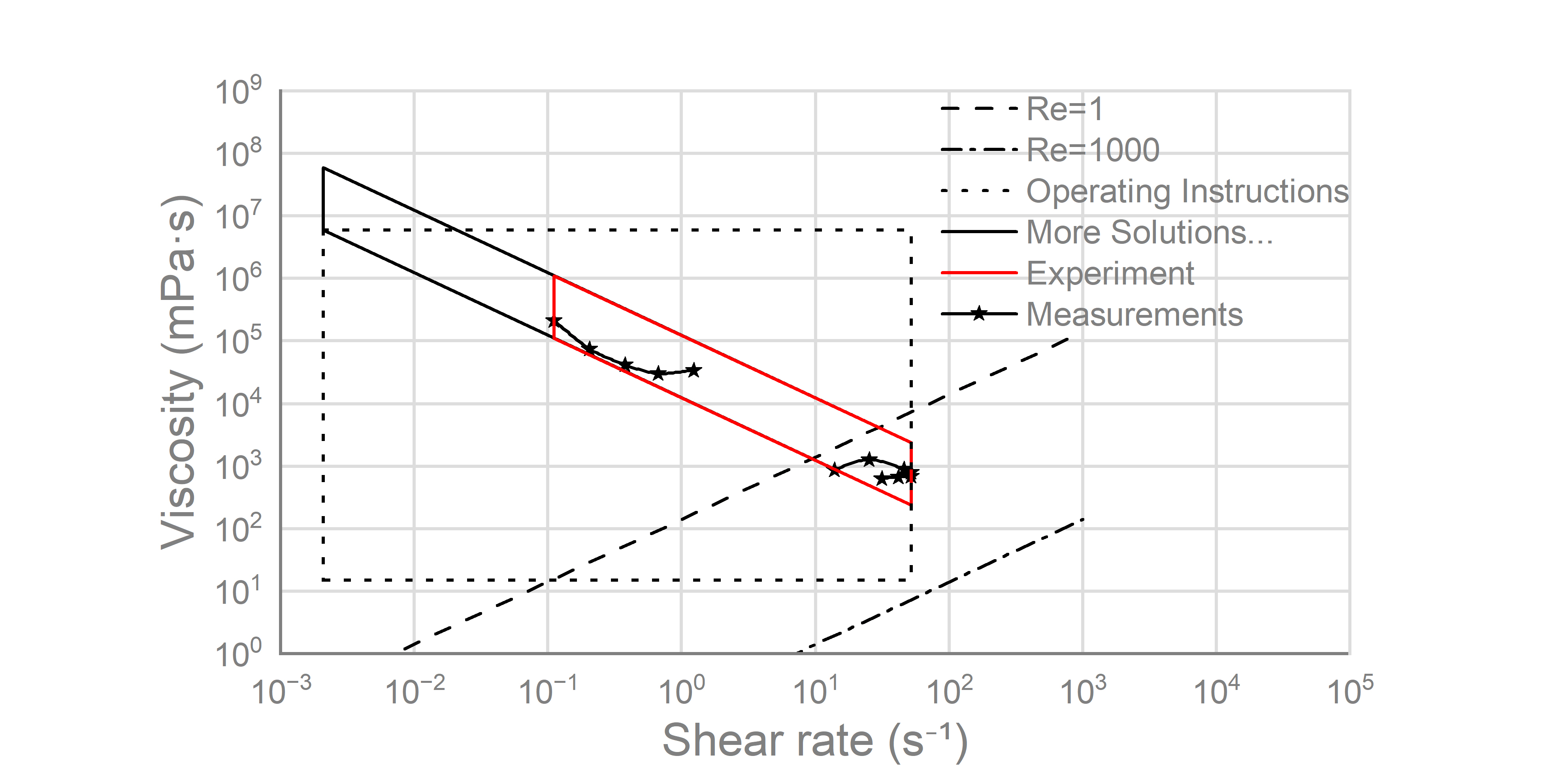}
\caption{Operating range of the Brookfield viscometer for the LV4 spindle in viscosity–shear rate coordinates. The dashed lines correspond to Reynolds numbers $\mathrm{Re}=1$ and $\mathrm{Re}=1000$. The dotted rectangle indicates the manufacturer-recommended operating region. The solid lines represent theoretical viscosity limits. Experimental data are shown by symbols, and the red polygon highlights the range covered in the present study.}
\label{fig:lv4_operating_range}
\end{figure}

{Figures ~\ref{fig:ul_adapter_operating_range}--\ref{fig:lv4_operating_range} (UL Adapter, LV1–LV4) show the applicability ranges of the corresponding geometries of the Brookfield viscometer in viscosity–shear rate coordinates plotted on a logarithmic scale. The applicability region specified in the \cite{bib2} technical documentation is outlined by a dashed black boundary, while the region reported in \cite{bib3} is shown by a solid black boundary; the solid red boundary marks the region actually covered by the experimental measurements in the present study. The dashed and dash–dotted lines indicate the theoretical boundaries corresponding to the Reynolds numbers $Re$ = 1 and $Re$ = 1000, respectively, calculated for a fluid with water density of 20~\si{\degreeCelsius}, taking into account the geometric parameters of each measuring system. For all measuring systems, the experimentally covered range lies entirely within the region specified in the \cite{bib3} technical documentation, where the calculations are based on the upper permissible values of viscosity and rotor rotation speed.}

{For the UL Adapter system, the experimentally covered region is divided by the $Re$ = 1 line into a laminar flow zone and a zone corresponding to the onset of flow disturbances and end effects. At the same time, the entire experimental region lies well below the $Re$ = 1000 line, indicating the absence of a turbulent flow regime over the entire measurement range (Fig.~\ref{fig:ul_adapter_operating_range}). Therefore, edge effects for the UL Adapter system can be neglected under experimental conditions, and measurements may be regarded as hydrodynamically valid.}

{For the LV1 spindle, the experimentally covered region likewise occupies only a small fraction of the applicability range specified in the technical documentation (Fig.~\ref{fig:lv1_operating_range}). The $Re$ = 1 line intersects only a small portion of the diagram, whereas the $Re$ = 1000 line lies well below the experimental range, confirming the absence of a turbulent flow regime under the experimental conditions. Consequently, viscosity measurements performed with the LV1 spindle within the investigated range are not constrained by Reynolds number limitations.}

{A similar situation is observed for the LV2 spindle (Fig.~\ref{fig:lv2_operating_range}). The experimentally achievable range is considerably narrower than the range specified in the technical documentation, yet both lie entirely within the $Re$ $<$ 1 region. The $Re$ = 1 boundary does not intersect the technical applicability region, indicating that flow turbulence does not occur when using the LV2 spindle for liquids with viscosities higher than that of water.}

{For the LV3 spindle, the experimentally covered region is shifted toward higher viscosity values relative to LV1–LV2 due to the geometric characteristics of the measuring system (Fig.~\ref{fig:lv3_operating_range}). The experimental region is divided by the $Re$ = 1 line into a laminar-flow zone and a zone where the onset of flow disturbances and end effects may occur. However, the entire experimental region lies well above the $Re$ = 1000 line, indicating that the flow regime remained laminar throughout the measurements.}

{For the LV4 spindle, which has the largest radial gap, only a small part of the applicability region defined in the technical documentation was experimentally covered (Fig.~\ref{fig:lv4_operating_range}). In addition, the $Re$ = 1 and $Re$ = 1000 lines are located closer to the experimental region compared with the other spindles of the LV series. Nevertheless, even in this case, the experimentally covered range lies almost entirely within the $Re$ $<$ 1 region.  This indicates that, for highly viscous liquids within a limited shear-rate range, the use of the LV4 spindle does not result in a transition to turbulent flow.}

\begin{figure}[H]
\centering
\includegraphics[width=\textwidth]{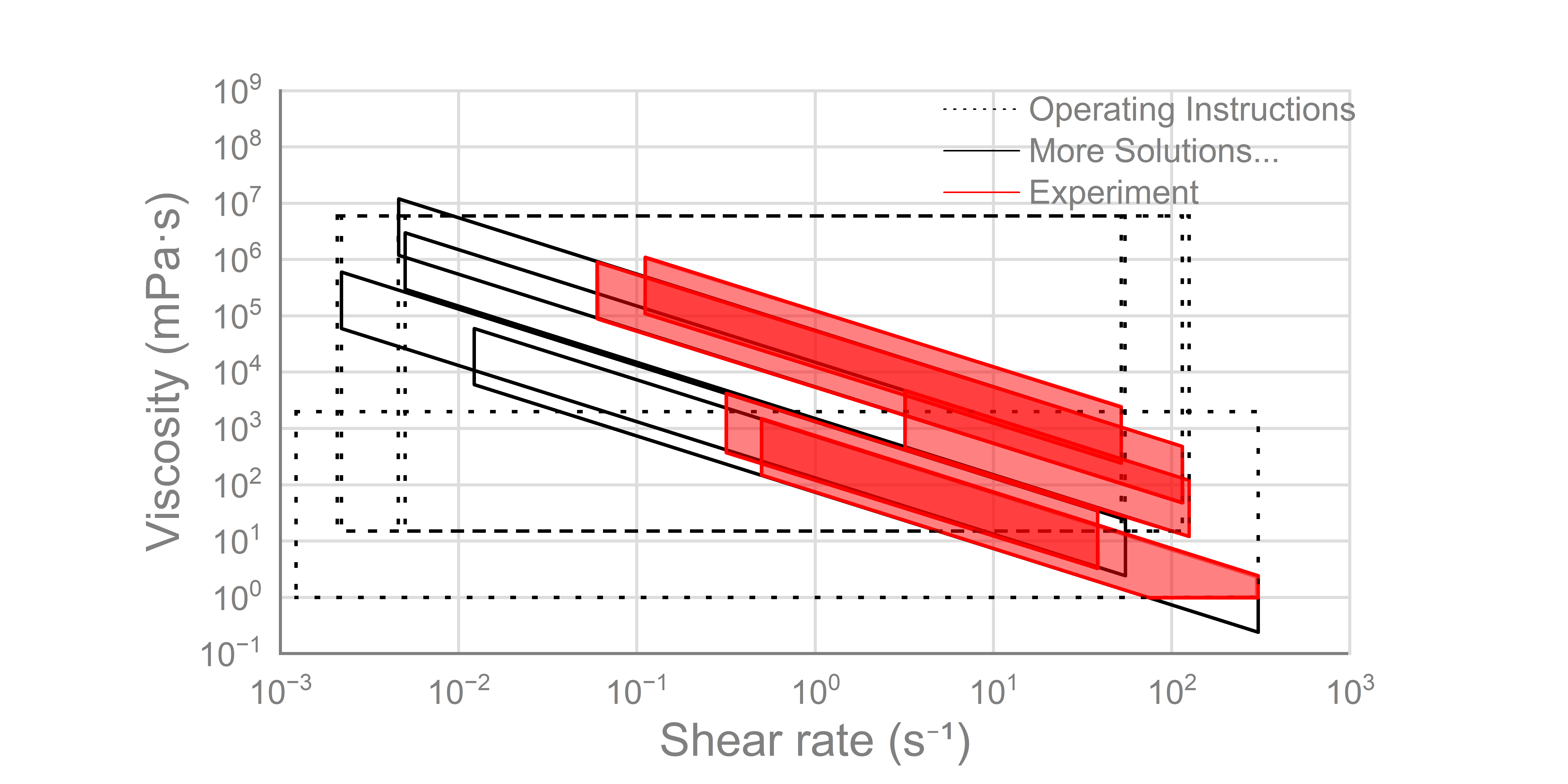}
\caption{Combined operating ranges of the Brookfield viscometer for all investigated spindles in viscosity–shear rate coordinates. The dotted rectangles indicate the manufacturer-recommended operating regions. The solid lines represent theoretical viscosity limits for different geometries. The red shaded regions correspond to the experimental domains covered in the present study for each spindle.}
\label{fig:combined_operating_range}
\end{figure}

{The analysis showed that, for all considered measuring systems of the Brookfield viscometer (UL Adapter, LV1–LV4), the experimentally realized viscosity and shear-rate ranges are narrower than those specified in the manufacturer’s technical documentation (Fig.~\ref{fig:combined_operating_range}). One of the main reasons for this discrepancy is the limitation of the operational torque range. Data obtained at torque values below 10\% and above 90\% of the instrument’s nominal range were excluded in accordance with the manufacturer’s recommendations to ensure measurement reliability and reproducibility \citep{bib2,bib3}. Additional factors that reduced the experimental region included limitations related to flow stability (particularly for cross-linked gels) and practical limits on achievable rotational speeds for specific spindle–fluid combinations.}

{The theoretical boundaries corresponding to $Re$ = 1 and $Re$ = 1000 were constructed to assess the potential influence of inertial effects and the onset of unsteady flow regimes. Water at 20~\si{\degreeCelsius} was used as the reference fluid for the calculations because it has the lowest viscosity among the studied systems and therefore the greatest tendency to produce higher Reynolds numbers under otherwise identical conditions. Thus, the performed assessment is conservative in nature and establishes the upper limit of applicability of the measuring systems in terms of hydrodynamic conditions. Analysis of the position of the experimental regions relative to the $Re$ = 1 and $Re$ = 1000 lines showed that, in all cases, the measurements were carried out in a flow regime far from turbulent conditions. This indicates that applicability limitations related to the possible development of edge and end effects did not significantly influence the results of the present study, and that the obtained experimental data are valid under the hydrodynamic conditions realized during the measurements.}

\subsection{Low-shear measurements}\label{subsec2}

\begin{figure}[H]
\centering
\includegraphics[width=\textwidth]{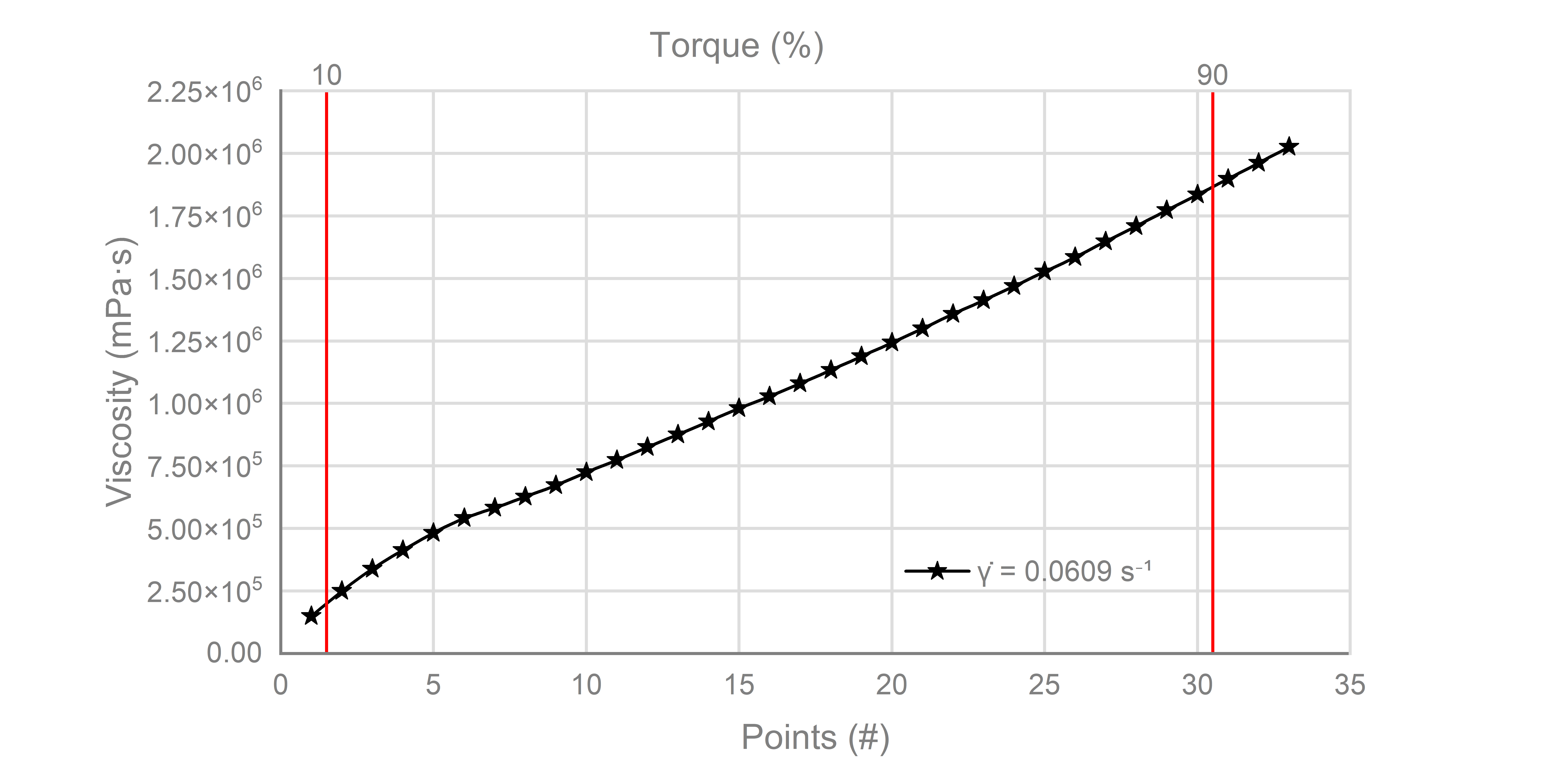}
\caption{Time evolution of viscosity for a cross-linked guar gel at a constant shear rate $\dot{\gamma} = 0.0609\ \mathrm{s^{-1}}$. The vertical red lines indicate the recommended torque range (10–90\%). Despite a measurement duration significantly exceeding the criterion $t \geq 1/\dot{\gamma}$, the viscosity does not reach a steady-state plateau and continuously increases, spanning the entire allowable torque range of the instrument.}
\label{fig:guar_time_dependence}
\end{figure}

{Despite the recommendation of \cite{bib7} that the measurement time for each viscosity point should satisfy t $\geq$ ${\dot{\gamma}^{-1}}$, stable viscosity data for the nonlinear cross-linked guar gel could not be obtained even after more than 30 minutes of testing at a constant shear rate ($\dot{\gamma}$ = 0.0609 \si{s^{-1}}). Based on this recommendation, the measurement of a single point should take at least $\approx$16.5 \si{s}); however, the actual measurement time exceeded this value by more than 100 times (1960 \si{s}). During this time, the viscosity curve did not reach a plateau and passed through the entire permissible torque range of the instrument (Fig.~\ref{fig:guar_time_dependence}). Therefore, when measuring highly viscous gels at low shear rates, particular care should be taken to ensure that the viscosity curve reaches a plateau at each shear rate.}

\section{Discussion}\label{sec5}

{From a fluid mechanics perspective, obtained results can be interpreted in terms of velocity gradient distribution in confined rotational flows, where the characteristic shear rate is governed by geometry-dependent hydrodynamic conditions. In addition, the results obtained in this study provide a consistent framework for the quantitative interpretation of Brookfield DV3TLV measurements, combining experimentally determined correction coefficients with physically justified shear rate calculations and flow regime analysis.}

{In particular, the derived conversion coefficients for the LV2 and LV3 spindles demonstrate that the relationship between rotational speed and shear rate is primarily governed by spindle geometry and remains stable across a wide range of fluids, including both Newtonian and non-Newtonian systems.}

{Based on a direct comparison with Anton Paar measurements, it can be concluded that when the actual spindle geometry is taken into account and the rotational speed is converted to shear rate, the Brookfield viscometer can adequately describe the rheological behavior of fluids in the medium and high viscosity ranges. However, when measuring Newtonian fluids, the accuracy of the instrument is limited, which is consistent with the observations of \cite{bib7} regarding the insufficient sensitivity of coaxial viscometers at low shear stresses.}

{Special attention was given to determining the flow regime during rotational flow measurements. Calculation of the Reynolds number according to the approach described by \cite{bib7} and \cite{bib5} showed that for most non-Newtonian fluids the flow remained laminar ($Re$ $<$ 1), confirming the applicability of a quasi-static interpretation. However, when using the UL Adapter and at high shear rates, conditions close to the transitional regime were observed.}

{One of the key outcomes of this study was the derivation of analytical formulas for converting rotational speed into shear rate while accounting for the actual spindle geometry. These equations allow accurate coefficients to be obtained (0.50 for LV2 and 0.46 for LV3), comparable to the geometrically derived values reported in the classical works of \cite{bib7} and \cite{bib10}. They provide a more accurate determination of shear rate than the standard approximate expression.}

{Therefore, Brookfield viscometer can be used for a preliminary quantitative evaluation of rheological properties when high accuracy is not required, provided that several conditions are satisfied. These include the use of recalculated shear rates based on geometric relations, control of the Reynolds number to confirm a laminar flow regime, restriction of the instrument’s application to medium- and high-viscosity fluids with pronounced non-Newtonian behavior, and comparison of the obtained results with rheometer data while accounting for the corresponding shear stress values.}

{The developed methodology is relevant for rapid engineering analysis, quality control of fluids in the oil and chemical industries, and laboratory evaluation of formulation modification effects. Further research may focus on accounting for the influence of spindle end surfaces, refining the stress gradient, modeling the viscosity distribution in the gap, and considering temperature effects \citep{bib8,bib12}.}

{The obtained results provide a basis for applying the Brookfield DV3TLV viscometer not only to classical fluids but also to more complex systems, such as nanomodified polymers, where viscosity control under high temperature and salinity conditions is required. \cite{bib4} demonstrated that the addition of metal oxide nanoparticles significantly increases the stability and viscosity of xanthan solutions; however, such experiments were performed using high-end rheometers. The availability of correction coefficients and conversion relationships proposed in the present study makes it possible to perform similar investigations using the Brookfield DV3TLV with accuracy sufficient for engineering applications.}

{Unlike previous studies, which primarily focused on empirical calibration or specific fluid systems, the present work integrates geometric analysis, experimental validation, and flow regime assessment into a single methodology. This approach enables not only improved accuracy of rotational flow measurements but also a clear definition of the physical limits of applicability of Brookfield-type viscometers.}

\section{Conclusion}\label{sec5}

{The results demonstrate that flow behavior in rotational viscometry systems is primarily governed by geometric and hydrodynamic factors. In this study, a comprehensive approach to the interpretation of viscosity measurements obtained with the Brookfield DV3TLV viscometer has been developed, combining geometric analysis, experimental validation, and flow regime assessment.}

{It is demonstrated that, when the actual spindle geometry is taken into account, the relationship between rotational speed and shear rate can be reliably described using constant conversion coefficients for the LV2 and LV3 spindles. The obtained values, $k = 0.50124 \pm 0.00085~\mathrm{s^{-1}\cdot rpm^{-1}}$ and $k = 0.46019 \pm 0.00120~\mathrm{s^{-1}\cdot rpm^{-1}}$, demonstrate high stability across a wide range of Newtonian and non-Newtonian fluids, confirming their predominantly geometric origin.}

{Comparison with measurements performed on the Anton Paar MCR 302 rheometer showed that the use of correction coefficients allows the apparent viscosity obtained with the Brookfield viscometer to be converted into values quantitatively consistent with reference data. Under controlled conditions (torque range $10$--$90\,\%$ and laminar or near-laminar flow regime), the deviation does not exceed a few percent for medium- and high-viscosity fluids.}

{The analysis of the Reynolds number demonstrated that, for most investigated systems, the flow remains within the laminar regime, which justifies the application of quasi-static models. At the same time, for low-viscosity fluids and high shear rates, transitional regimes and end effects become significant and limit the reliability of rotational flow measurements.}

{Hence, the results define the physical limits of applicability of the Brookfield DV3TLV viscometer and provide a consistent framework for interpreting its measurements in terms of absolute rheological properties.}

{From a practical perspective, the proposed methodology can be used for rapid quantitative evaluation of viscosity in industrial and laboratory conditions where high-end rheometers are unavailable, particularly in the oil and chemical industries. The use of the UL Adapter is recommended for low-viscosity fluids, whereas the LV2 and LV3 spindles provide the most reliable results for medium- and high-viscosity systems when combined with the proposed conversion and correction procedures.}

\subsection{Practical recommendations}\label{subsec2}

{The practical use of the Brookfield DV3TLV viscometer can be summarized as follows. When interpreting measurement results obtained with the LV2 and LV3 spindles, the universal conversion coefficients from rotational speed to shear rate determined in this study should be applied, namely $k = 0.50124 \pm 0.00085~\mathrm{s^{-1}\cdot rpm^{-1}}$ and $k = 0.46019 \pm 0.00120~\mathrm{s^{-1}\cdot rpm^{-1}}$, respectively (Table~\ref{tab:k_LV2}, \ref{tab:k_LV3}).}

{Comparison with the data obtained on the Anton Paar MCR 302 rheometer showed that, when the torque range of 10--90\% is maintained and the Reynolds number is controlled, the Brookfield DV3TLV viscometer provides quantitatively comparable results for medium- and high-viscosity fluids. For a 92.5\% glycerol solution at 20~\si{\degreeCelsius}, the best agreement was achieved using the LV3 spindle and the UL Adapter system Table~\ref{tab:kcorr_all}. For 50\% glycerol and fluids with viscosities on the order of several \si{\milli\pascal\second}, the applicability is limited to the UL Adapter and a narrow shear rate range due to the increase in Reynolds number and the influence of end effects. In the case of linear guar gels, the UL Adapter provides the best reproducibility and minimal deviations, whereas for cross-linked gels at 70~\si{\degreeCelsius} the use of LV1 and the UL Adapter is acceptable, while the application of LV2 under these conditions is associated with significant systematic errors Table~\ref{tab:kcorr_all}.}

\section{Data availability}\label{sec5}

{The datasets generated and analyzed during the current study are available from the corresponding author on reasonable request. Relevant data supporting the findings of this study are included within the article.}

\section{Competing interests}\label{sec5}

{The authors declare no potential conflicts of interest with respect to the research, authorship, and/or publication of this article.}

\bibliography{sn-bibliography}

\end{document}